\documentclass[fleqn,usenatbib]{mnras}
\usepackage{newtxtext,newtxmath}
\usepackage[T1]{fontenc}
\DeclareRobustCommand{\VAN}[3]{#2}
\let\VANthebibliography\thebibliography
\def\thebibliography{\DeclareRobustCommand{\VAN}[3]{##3}\VANthebibliography}
\usepackage{graphicx}	
\usepackage{amsmath}	
\usepackage{siunitx}
\usepackage{relsize}
\usepackage{comment}
\usepackage{adjustbox}

\usepackage{multirow}

\DeclareMathOperator{\sign}{sign}
\title[Effect of stellar spots on ARIEL observations]{Correcting the effect of stellar spots on ARIEL transmission spectra II. The limb darkening effect}

\author[G. Cracchiolo et al.]{
G. Cracchiolo,$^{1,2}$\thanks{E-mail: gianluca.cracchiolo@inaf.it}
G. Micela,$^{1}$
G. Morello$^{1,3,4}$
and G. Peres$^{1,2}$
\\
$^{1}$INAF- Palermo Astronomical Observatory, Piazza del Parlamento, 1, 90134 Palermo, Italy \\
$^{2}$University of Palermo, Department of Physics and Chemistry “Emilio Segrè”\\
$^{3}$ Departamento de Astrof\'isica, Universidad de La Laguna (ULL), 38206, La Laguna, Tenerife, Spain\\
$^{4}$ Instituto de Astrof\'isica de Canarias (IAC), 38205 La Laguna, Tenerife, Spain
}
\date{Accepted 2021 August 10. Received 2021 July 14; in original form 2021 March 31}
\pubyear{2020}

\begin{document}
\label{firstpage}
\pagerange{\pageref{firstpage}--\pageref{lastpage}}
\maketitle

\begin{abstract}
    This paper is part of an effort to correct the transmission spectra of a transiting planet orbiting an active star. In Paper I (\citealt{Cracchiolo}) we have demonstrated a methodology to minimize the potential bias induced by unocculted star spots on the transmission spectrum, assuming a spot model parameterized by filling factor and temperature. In this work we introduce the limb darkening effect, therefore the position of the spot in the stellar disk and the impact parameter of the transiting planet now play a key role. The method is tested on simulations of planetary transits of three representative kinds of planetary systems, at ARIEL resolution.
    We find that a realistic treatment of the limb darkening is required to reliably estimate both the spots parameters and the transmission spectrum of the transiting planet.
    Furthermore, we show that the influence of the spots on the retrieval of the planetary transmission spectrum is significant for spots close to the center of the star, covering a fraction greater than $0.05$ and with a temperature contrast greater than $500\,K$, and that for these cases our method can confidently extract the transmission spectrum and the impact parameter of the transiting planet for both cases of occulted and not occulted spots, provided that we have an accurate characterization of the stellar parameters  and a reliable simulator of the instrument performances.
\end{abstract}




\section{Introduction}\label{introduction}

    This paper is part of a project for the extraction of the transmission spectrum of a transiting planet in the presence of stellar activity-related phenomena on the host star. The project is motivated by the ARIEL space mission (\citealt{ARIEL}) whose goal is the observation of the atmospheres of about $1\,000$ known extrasolar planets, in a broad wavelength range from the visible to the infrared (0.5-8 \si{\micro\metre}). The project starts from \cite{Cracchiolo} (henceforth Paper I), in which we have presented a basic model of the stellar activity dominated by the presence of spots, by neglecting other photospheric inhomogeneities, or the limb darkening effect. The idea behind the project is to introduce, step by step, so as to ascertain carefully their effect, other components in our stellar activity model and, in this paper, we will study the role of the limb darkening effect.\\
    
    In Paper I, the assumption of a uniform emission from both the spot and the quiet photosphere has allowed us to define the spots in terms of two parameters, i.e. the fraction of the spotted stellar disk (filling factor) and the spot temperature. We have found that it is possible to model the stellar activity due to the spots from the out-of transit observation in terms of these two parameters. In particular, for a given target, we have built a 2-d grid of spotted spectra where each element has an assigned combination of filling factor and spot temperature, then we compared the out-of transit spectra with this grid, in order to derive the combination of the spot parameters that better match the observations. This info is used to recover the planet transmission spectrum during the transit, in the hypothesis of a stellar disk with a uniform emission. With this assumption the transit depth depends uniquely on the planet-to-star areas ratio and not on other orbital parameters of the transiting system. We have found that if the spot filling factor is greater than $0.02$ and the spot temperature is $300\,K$ cooler than the photospheric temperature it is necessary to correct the spectrum in order  to reliably extract the transmission spectrum of the transiting planet. However, ignoring the limb darkening effect may limit our capability to retrieve the spot parameters and  correctly derive the planetary transmission spectrum, as we will discuss in this  paper. \\
    
    Here we introduce the limb darkening effect, therefore the star has a non uniform emission, and both the position of the spot and the planet impact parameter become crucial parameters. In continuity with what done in Paper I, we adopt a stellar activity model with one dominant spot on the visible stellar hemisphere, but, in addition to its filling factor and its temperature, the spot is characterized by its position on the stellar disk, here parameterized in terms of the distance between the spot and the center of the disk. Therefore, the spot is now characterized by three parameters (no longer two). In analogy with Paper I, we present an approach to retrieve the spot parameters from the out-of transit observation, by comparing the out-of transit observation of a star with a 3-d grid of stellar spectra specifically generated for that target, where each element of the grid models the star with a specific spot on the disk and different elements correspond to different filling factors, temperatures and distances from the stellar disk center. 
    
    The limb darkening effect changes the shape of the primary transit light curve, as the planet occults portions of the stellar disk with a non-uniform emission, therefore the transit profile is determined not only by the planetary radius (as assumed in Paper I), but also by the limb darkening profile of the star, as well as by the orbital parameters of the transiting system. For this reason, here we need to simulate realistic light curves of transiting planets. The presence of spots and, more in general, of stellar activity phenomena, induces chromatic perturbations of the transit profile (e.g., \citealt{Sing_2011,Pont_2013,Rackham}), therefore it may affect not only the determination of the planetary radius with a wavelength-dependent effect but also the derivation of the orbital parameters and the limb darkening coefficients. Most works consider only the planetary radius variation (e.g., \citealt{Berta_Hubble,Carter}), but the effect due to the spots on the orbital parameters may be significant (see, e.g., \citealt{Ballerini}). 
    
    The effect of star spots is to overestimate the transit depth if they are not crossed during the transit, and to underestimate the transit depth if there are spot crossings, with a chromatic dependence (e.g., \citealt{Sing_2011, Ballerini, Pont_2013, McCullough, Zellem}). Here, we quantify the effect of the spots on the derivation of the transmission spectrum and of the impact parameter of a transiting planet, in the assumption that all the other orbital parameters of the transiting system are known and fixed (see Section \ref{retrieval_planet}). We also present a method to retrieve both the planet transmission spectrum and its impact parameter in the presence of spots, that consists in modeling the flux from the quiet photosphere blocked by the transiting planet. The approach is based on the assumption that the spot evolves on a time scale much longer than the transit duration (as discussed in Paper I), so that, if the spot is not crossed during the transit, the spot influence may be removed when subtracting the out-of transit stellar flux from the entire transit observation. The method allows to retrieve the planet transmission spectrum and its impact parameter: the latter was not determined in Paper I as the transit profile was assumed to depend only on the ratio between the planet and the star areas at each wavelength.\\
    
    Similarly to what was done in Paper I, we work at ARIEL Tier 2 resolutions since ARIEL Tier 2 is the largest ARIEL sample for which we will obtain quantitative measurements of the atmospheric properties (see \citealt{Tinetti_2018} for a description of ARIEL's tiering approach).
    Tab. \ref{tier} shows the Tier 2 resolutions for each spectral channel of ARIEL.
    The approach is tested on realistic simulations of ARIEL observations of the three targets of planetary systems simulated in Paper I (HD 17156 b, HAT-P-11 b, K2-21 b), chosen to sample a range of exoplanetary and stellar conditions and simulated with the Ariel radiometric model (\citealt{ArielRad}).

    \begin{table}
        \centering
        \caption{ Spectral Resolution (R) of the ARIEL Tier 2 data for each spectral channel. $\Delta \lambda$ is the wavelength coverage of each ARIEL channel.}
        \begin{adjustbox}{width=0.7\columnwidth}
        \begin{tabular}[h!]{lcc}
        \hline
        Channel     & $\Delta \lambda$ & Tier 2 \\
                       &     (\si{\micro\meter})    & (R)    \\
        \hline
        VISPhot        &     $0.5-0.6$    &   1       \\
        FGSPrime       &     $0.6-0.8$    &   1       \\
        FGSRed         &     $0.8-1.1$    &   1       \\
        NIRSpec        &     $1.1-1.9$    &  $\sim$10 \\
        AIRS-CH0       &     $1.9-3.9$    &  $\sim$50 \\
        AIRS-CH1       &     $3.9-8.0$    &  $\sim$10 \\
        \hline
        \end{tabular}
        \end{adjustbox}
        \label{tier}
    \end{table}
    
    In Section \ref{models} we present our models of active stars and the tools used to simulate the transit light curves of the three planetary systems and their atmospheres. In Section \ref{method} we describe our method to detect the presence of spots from the out-of transit observation and then extract the planetary transmission spectrum from the in-transit observation. Section \ref{results} shows our results, while Section \ref{conclusion} contains a summary of our work and its possible applications. Appendices \ref{spot_fraction} and \ref{intersection_fraction} contain some mathematical derivations used for our simulations, while Appendix \ref{Discussion} shows the results obtained when applying the method without taking into account the limb darkening (Paper I) to data simulated in the presence of limb darkening effect.

\section{Models and simulations}\label{models}
    This Section contains a detailed description of the modeling of transit observations in the presence of stellar spots and the simulations of three potential ARIEL targets taken from the catalogue of \cite{Billy}: HD 17156 b, a hot Jupiter orbiting a G-type star (\citealt{Discovery_jupiter}), HAT-P-11 b, a Neptune-size planet orbiting a K star ( \citealt{Discovery_neptune}) and K2-21 b, a super-Earth orbiting a M type star ( \citealt{rotation_K2_21}). We simulate planetary transits and the simultaneous presence, on the stellar disk, of spots having different sizes, temperatures and positions on the stellar disk in order to understand their impact on the retrieval of the planetary transmission spectrum.

\subsection{Stellar atmosphere models}
    We simulate the stellar atmospheres with the models in the BT-Settl library\footnote{\url{https://phoenix.ens-lyon.fr/Grids/BT-Settl/.}} (\citealt{Baraffe}), where each spectrum is identified by three stellar parameters: the effective stellar temperature ($T_{\rm eff}$), the stellar surface gravity ($\log{g}$), and the stellar metallicity ($[M/H]$). In this grid the stellar metallicity is set to the solar abundances ($[M/H]=0$), while the effective temperature spans the $1200-7000$ \si{\kelvin} interval, and the logarithm of the surface gravity ranges in the interval $[2.5-5.5]$. As in Paper I, we assume that the spectrum of the stellar spots is equivalent to a stellar spectrum of lower temperature, but with the same $\log{g}$\footnote{Some authors  (e.g. \citealt{Solanki}) model the stellar spots as having a 0.5-1 dex lower $\log{g}$ than the stellar one, and this is due to the decrease in gas pressure caused by increased magnetic pressure in the darker photospheric upper layers.}. Here we use models with $\log{g} = 4.5$, i.e. with Sun-like surface gravity, for both the spots and the photosphere. The BT-Settl models are not spatially resolved and give the stellar flux integrated over the entire surface. However in this work we want to assess the impact of the limb darkening, which is the new aspect relative to the Paper I. In order to approximate the intensity profiles we use the four-coefficients law proposed by \cite{Claret}, since it is currently the  most precise analytical representation of the limb darkening ( \citealt{2011MNRAS.418.1165H,2017AJ....154..111M}):
    \begin{equation}
        \frac{I_{\lambda} (\mu)}{I_{\lambda} (1)} = 1-\sum_{n=1}^4 a_{n,\lambda}(1-\mu^{n/2})
        \label{ld_profile}
    \end{equation}
    where: 
    \begin{itemize}
        \item [--] $\lambda$ indicates a specific spectral bin/passband;
        \item [--] $\mu=\cos{\theta}$, being $\theta$ the angle between the line of sight and the normal inward direct the stellar surface;
        \item [--] $I_{\lambda} (\mu)$ is the stellar intensity profile and $I_{\lambda} (1)$ is the intensity at the center of the disk ($\mu=1$);
        \item [--] $a_{n,\lambda}$ are the limb darkening coefficients (LDCs).
    \end{itemize}
    Eq. \eqref{ld_profile} describes the intensity profile as a function of the variable $\mu$, which is related to the radial coordinate in the projected stellar disk by the relation $r=\sqrt{1- \mu^2}$, being $r=0$ at the center and $r=1$ the stellar disk radius. We adopt Eq. \eqref{ld_profile} to describe the intensity at a given radius on the disk for both the spots and the photosphere.
    We use the database \texttt{PHOENIX\textsubscript{-}2012\textsubscript{-}13} (\citealt{Claret_2012,Claret_2013}) of the \texttt{ExoTETHyS} package\footnote{\url{https://github.com/ucl-exoplanets/ExoTETHyS/}.} (\citealt{morello}) to derive the LDCs. Models in the \texttt{PHOENIX\textsubscript{-}2012\textsubscript{-}13} database are parameterized as the BT-Settl models: $T_{\rm eff}$ spans the range $3000-10\,000$ \si{\kelvin}, $\log{g}$ the range $[0.0-6.0]$ and, as above, $[M/H]=0$. We choose this library because it covers the ARIEL wavelength coverage and because it reaches low temperatures\footnote{A new grid \texttt{Phoenix\textsubscript{-}drift\textsubscript{-}2012} (\citealt{Claret_2012, Claret_2013}) has been recently developed which reaches down to 1500 \si{\kelvin}.}, useful for the simulation of cool spots.\\
    From the \texttt{ExoTETHyS} package we derive the normalized intensity profiles as in Eq. \eqref{ld_profile}; the absolute intensity profiles $I_\lambda (\mu)$ are derived by imposing Eq. \eqref{cons_flux}, i.e. that the stellar flux integrated over the stellar disk with the intensity profile in Eq.\eqref{ld_profile} is equivalent to the flux $F_\lambda$ coming from the BT-Settl models, both obtained with the same set of stellar parameters:
    
    \begin{equation}
        \label{cons_flux}
        F_\lambda=2\pi \int_0^1 I_\lambda(r) r dr = 2\pi \int_0^1 I_\lambda(\mu) \mu d\mu
    \end{equation}
    The last equality in the previous equation is obtained with the change of variable $r=\sqrt{1- \mu^2}$. If we replace in Eq. \eqref{cons_flux} the expression of $I_\lambda(\mu)$ in terms of the LDCs $a_{n,\lambda}$ of Eq. \eqref{ld_profile} we will obtain:
    \begin{equation}
        \label{abs_flux}
        I_\lambda (1) =\frac{F_\lambda}{2\pi(\frac{1}{2}-\frac{a_{1,\lambda}}{10}-\frac{a_{2,\lambda}}{6}-\frac{3a_{3,\lambda}}{14}-\frac{a_{4,\lambda}}{4})}
    \end{equation}
    By replacing $I_\lambda(1)$ in Eq. \eqref{ld_profile} with its expression in Eq. \eqref{abs_flux} we found the absolute intensity  profiles $I_\lambda(\mu)$.\\
    The spectra from the BT-Settl grid and the LDCs from the \texttt{PHOENIX\textsubscript{-}2012\textsubscript{-}13} models are downloaded for temperature values in the interval $3000-6100$ \si{\kelvin}, with a sampling step $\Delta T$ = 100 \si{\kelvin}. For temperatures that are not in this grid, we derive the stellar spectrum and the corresponding set of LDCs by linearly interpolating models in the BT-Settl and \texttt{PHOENIX\textsubscript{-}2012\textsubscript{-}13} grids with the closest temperatures.
    
\subsection{Simulating active stars}\label{Simulating active stars}
    The presence of spots on the stellar surface breaks the radial symmetry of the intensity profile $I_\lambda(r)$. We model the spots as circular structures (see, e.g., \citealt{Montalto}) at temperature $T_s$ lower than the photospheric temperature $T_*$. If the star has only one spot with radius $r_s$ (normalized to the stellar radius), we define as filling factor of the spot the ratio between the circular spotted area and the projected stellar disk ($ff=r_s^2$). The geometric projection of the spot on the stellar disk will be an ellipse, whose eccentricity depends on the fractional distance $d$ between the spot and the disk centers (see Appendix \ref{spot_fraction}). \\
    The out-of transit stellar flux $F_\lambda^{out}$ is obtained with a ring integration, i.e. dividing the stellar disk in $N$ equally spaced annuli and summing the contribution from each annulus as follows:
    \begin{equation}
        \label{model_out}
        F_\lambda^{out} = \sum_{i=1}^N I_\lambda (r_i)\cdot 2\pi r_i \cdot \Delta r
    \end{equation}
    where $r_i$ is the central radius of the $i^{th}$ annulus, $\Delta r$ is the distance between two neighbor annuli, and:
    \begin{equation}
        \label{spotted_profile}
        I_\lambda (r_i) = f(r_i)\cdot I_\lambda(r_i, T_s) +  \Big( 1-f(r_i) \Big)\cdot I_\lambda (r_i, T_*)
    \end{equation}
    The term $f(r_i)$ gives the fraction of the $i^{th}$ annulus covered by the spot, while  $I_\lambda(r_i, T_s)$ and $I_\lambda(r_i, T_*)$ are the darkened intensity values of the spot and of the star, respectively. In our analysis we assume the four-coefficients limb darkening law in Eq. \eqref{ld_profile} to simulate the intensity profiles of both the spot and the photosphere (a similar approach is used in \citealt{Csizmadia}).
    Each fraction $f(r_i)$ depends on the filling factor of the spot and its distance from the disk center and is, here, analytically computed (the detailed calculations are reported in Appendix \ref{spot_fraction}). The integrated stellar flux $F_\lambda^{out}$ is obtained by using $N=10\,000$ annuli. In the absence of spots, this choice allows to achieve an accuracy on the integrated flux $\lesssim 10^{-6}$ in a Sun-like star, at least in the ARIEL wavelength coverage. Achieving higher accuracies is not necessary for ARIEL as its noise floor, i.e. the lowest possible signal level that ARIEL may measure, is $20 \times 10^{-6}$.\\
    
    We can extend Eq. \eqref{spotted_profile} to the case of $M$ non-overlapping circular spots as:
    \begin{equation}
        \label{generalized_expression}
        I_\lambda (r_i) = \sum_{j=1}^M f^{j}(r_i)\cdot I_\lambda(r_i, T_s^{j})+  \Big( 1-\sum_{j=1}^M f^{j}(r_i) \Big)\cdot I_\lambda (r_i, T_*)
    \end{equation}
    where the $j$ index runs over the $M$ spots, each at temperature $T_s^{j}$.

\subsection{Simulating the planetary transmission spectrum}\label{pl_spectrum}
\begin{figure}
        \centering
        \includegraphics[width=0.9\columnwidth]{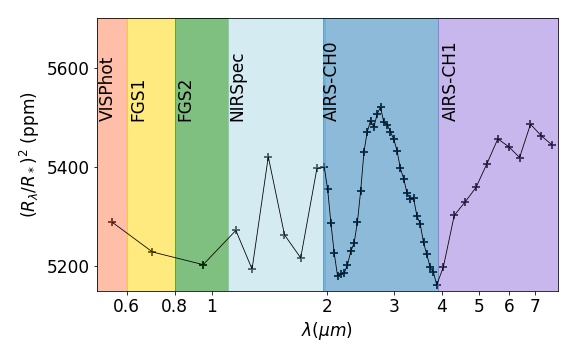}
        \caption{Synthetic transmission spectrum of the planet HD 17156 at ARIEL Tier 2 resolutions, in semi-log scale.}
        \label{transmission_spectrum}
    \end{figure}

    We use TauREx\footnote{\url{https://github.com/ucl-exoplanets/TauREx3_public}.}(\citealt{TauReX}) to simulate planetary transmission spectra. As in Paper I, we simulate primordial atmospheres, i.e. made of H$_{2}$ and He with an abundance ratio 0.17 (being the molecular hydrogen the main gas), and with traces of water (having volume mixing ratio $w=10^{-4}$). We adopt isothermal temperature profiles (the equilibrium temperatures $T_{eq}$ of the three considered planets are shown in Tab. \ref{parameters}) and we set the pressure $p_{max}$ at the bottom of the atmosphere to $1\, bar$ both for terrestrial and gaseous planets\footnote{Generally, for gaseous planets the pressure at the surface of the planet is $p_{0}\sim 1-10\, bar$, depending on the transparency of the atmosphere, \citealt{Tinetti_2013}}. In addition, we add the Rayleigh scattering of the molecular species in the planetary atmosphere to our model of transmission spectra. As an example, Fig.\ref{transmission_spectrum} shows the planetary transmission spectrum simulated for the planet HD 17156 b.
    
    \begin{table*}
    \centering 
    \caption{Parameters used to simulate the planetary transmission spectra and the transit light curves of our reference planetary systems. $(R_p/R_*)^2$, $M_p$, $T_{eq}$, $P$, $\frac{a}{R_*}$, $i$, the Transit Mid Time (in unit of Julian Date) and the K magnitude are taken from the catalogue of \protect\cite{Billy}. $b$ is the impact parameter of the transiting planet, derived from Eq. \eqref{impact_parameter}, $P_S$ is the stellar rotation period, and $n$ is the number of transits required to observe a primordial atmosphere on each target, if $S/N>7$.}
    \begin{adjustbox}{width=\textwidth}
    \begin{tabular}{c|ccc|cccccccc}
    \hline
    \hline
         & \multicolumn{3}{c|}{Transmission spectrum} &  \multicolumn{7}{c}{Transit light curve}\\
    \cline{2-12}
        Planet      &     $(\frac{R_p}{R_*})^2$    &   $M_p$         &  $T_{eq}$          &    $P$         &   $\frac{a}{R_*}$     &    $i$      & $b$    &  Transit Mid Time & K mag&    $P_S$ & $n$\\
    
                    &       (\%)                 &   (M$_J$)       &  (\si{\kelvin})            &   (\si{\day})     &                       &   (\si{\degree})  & &      (JD) & & (\si{\day})&\\
    
    \hline
    HD 17156 b    &      $0.5$                   &    $0.3$         &  $904$            &      $21.2$     &      $23.3$            &      $87.82$    & 0.89 &     $2458503.7$         & $6.8$ & $28.6^{(1)}$ & 2\\
    HAT-P-11 b     &      $0.33$                   &   $0.05$         &   $847.8$            &     $4.9$      &      $16.6$           &      $88.3$ & 0.5    &     $2458489.2$     & $7$ & $29.2^{(2)}$ & 3\\
    K2-21 b   &      $0.06$                  &   $0.01$        &   $647.2$            &     $9.3$     &      $25.2$           &      $88.98$    & 0.45 &     $2456989.4$    & $9.4$& $-$ & 184\\
    \hline
    \end{tabular}
    \end{adjustbox}
    {\raggedright (1) \citet{period_HD17156}; 
    (2) \citet{Bakos_2010}.\par}
    \label{parameters}
    \end{table*}

    \subsection{Light curves simulations}\label{simulate_light_curves}
    The transits observations are simulated with \texttt{pylightcurve 4.0.1}\footnote{\url{https://github.com/ucl-exoplanets/pylightcurve}.} (\citealt{pylightcurve}), a python package for modeling and analysing transit light curves. \texttt{pylightcurve} implements the analytical solution of \cite{Mandel} to compute the fractional dimming of the stellar flux $\delta_\lambda $ due to the transiting planet in a star without photospheric inhomogeneities. The package uses the orbital parameters of the transiting system (orbital period $P$, ratio $\frac{a}{R_*}$, being $a$ the semi-major orbital axis and $R_*$ the stellar radius, eccentricity $e$, inclination $i$, argument  of the periastron $\Omega$, and transit mid time), the planet to star radii ratio, and a well-defined limb darkening law. Here the orbital parameters of the three reference planetary systems are taken from the catalogue of \cite{Billy} and are shown in Tab. \ref{parameters}. We set the orbital eccentricity of the three planets to $e=0$. Since we want to simulate an atmosphere, we use a wavelength-dependent planet to star radii ratio (see Section \ref{pl_spectrum}). We describe the limb darkening profile with the Claret-4 law in Eq. \eqref{ld_profile}. 
    
    In the presence of a spot on the stellar disk, the in-transit stellar flux $F_\lambda^{in}$ is expressed as: 
    \begin{equation}
        \label{model_in}
        F_\lambda^{in}= F_\lambda^{out}-\delta_\lambda (t) \cdot P_\lambda (T_*) +\Gamma_\lambda (T_*,t)-\Gamma_\lambda (T_s,t) 
    \end{equation}
    where the first term is the out-of transit observed flux from Eq. \eqref{model_out}, $\delta_\lambda (t)$ is the fraction of non spotted flux occulted by the planet at a time $t$, $P_\lambda (T_*)$ is the photospheric flux integrated over the whole stellar disk, while the terms $\Gamma_\lambda (T_*,t)$ and $\Gamma_\lambda (T_S,t)$ are the fluxes of the photosphere and of the spot coming from the region where the spot and the planet overlap:

    \begin{equation}
        \Gamma_\lambda (T,t)=\sum_{i=1} ^N c_\lambda (r_i,t) \cdot I_\lambda (r_i,T)\cdot 2\pi r_i \cdot \Delta r
    \end{equation}
    being $c_\lambda (r_i,t)$ the fraction of the $i^{th}$ annulus intercepted by the occulted spot at time $t$. 
    The fraction $c_\lambda(r_i,t)$ is computed analytically in the hypothesis of circular spots with elliptical projections (see Appendix \ref{intersection_fraction}). Eq. \eqref{model_in} can be easily generalized to the case of $M$ non-overlapping circular spots; in this case $F_\lambda^{out}$ is generalized as described in Section \ref{Simulating active stars}, while the last two terms refer to the crossed spot.\\
    Eq. \eqref{model_in} assumes that the spot pattern during the transit observation does not change and this is a reasonable assumption for the stars in our sample, since the spot variability is modulated on a time scale comparable to the stellar rotation period $P_S$, that is much longer than the transit duration (see Tab. \ref{parameters}). Under this assumption, the first term $F_\lambda^{out}$ in Eq. \eqref{model_in} does not vary during the transit observation. If instead the star rotated quickly, the spot would move during the transit, causing significant variations of the observed stellar flux due to the spots modulation (e.g. \citealt{CoRoT, Ioannidis}). \\
    
    Here, the light curves are simulated with a frequency of 1 data point every $100$ \si{\second} in all ARIEL channels, by assuming that the total transit observation is 2.5 times the transit duration $T_{14}$\footnote{We follow the common nomenclature to indicate the time interval between the first and fourth contact of the stellar and planetary silhouettes as $T_{14}$ (\citealt{Seager}).}   (\href{https://sci.esa.int/documents/34022/36216/Ariel_Definition_Study_Report_2020.pdf}{ESA/SCI(2020)1}).

\subsection{ARIEL Radiometric model}\label{ArielRad}
    As in Paper I, we use the ARIEL Radiometric Model, or ArielRad (\citealt{ArielRad}), to have realistic simulations of the ARIEL transit observations. ArielRad reads input spectra that are spatially integrated in a grid of spectral bins ranging the interval $0.4-8.5$ \si{\micro\meter} and simulates the stellar signal in the ARIEL photometric and low resolution spectral channels.  The integrated out-of transit and in-transit stellar fluxes in each of these spectral bins are obtained with an annuli integration according to Eqs.\eqref{model_out} and \eqref{model_in}, respectively. Here, we use the ArielRad version 17r8.\\
    The noise level associated to each simulated spectrum follows the equation (5) of Paper I. We emphasize that the ArielRad output noise is dominated by the photon noise, therefore it mainly depends on the magnitude of each simulated star (the K mag of each star is reported in Tab. \ref{parameters}) and on the exposure time of the simulation.
    
\section{Methods}\label{method}
    In this Section we assume a model of stellar activity with one dominant circular spot and we present a method to characterize the geometrical and physical properties of this “effective” spot from the out-of transit observation taking into account for limb darkening. In addition, we quantify the effect of the star spots in the extraction of a planet transmission spectrum and we present a method to correctly derive the transmission spectrum for transits in the presence of star spots.\\
    
    The approach in this paper is an upgrade of that used in the Paper I, described briefly below, where we assumed a stellar activity model dominated by star spots, on the hypothesis of a uniform emission from both the star and the spot. In Paper I, the spots are simulated as pieces of the photosphere at lower temperature and covering a fraction of the stellar disk, so they are characterized by two parameters: temperature and projected filling factor. The latter are retrieved by comparing the observed out-of-transit stellar flux with a 2-d grid of stellar spectra, specifically built for the observed target, where each spectrum corresponds to a specific combination of the two spot parameters. These parameters are used to recover the transmission spectrum of the planet from the in-transit observation, on the assumption that the transit depth is uniquely dependent on the planet-to-star area ratio at each wavelength. The latter assumption is valid only if the limb darkening effect is negligible.
    In this paper we add to our model the limb darkening effect that, as we will show, cannot be neglected and, therefore, in the following, the star is not modeled as a uniform disk, and both the distance $d$ of the spot from the center of the stellar disk and the impact parameter $b$ of the transiting planet become important. The main steps in this work are:

    \begin{enumerate}
        \item creating a 3-d grid of out-of-transit stellar spectra with spots having different temperatures, filling factors and positions on the stellar disk;
        \item  fitting the observed out-of-transit spectrum over this grid to retrieve the spot parameters;
        \item subtracting the average out-of-transit spectrum from the observation, thus obtaining the spectra from the regions of the stellar surface obscured by the planet;
        \item fitting the geometric parameters ($(R_p/R_*)^2$ and the planetary impact parameter $b$) that best reproduce the resulting time series of difference spectra. We also take into account the spots occulted by the planet, as detailed in Section \ref{retrieval_planet}.
    \end{enumerate}
 
 \subsection{Retrieval of the spot parameters}\label{retrieval_params}
    For each reference star, we build a 3-d grid of out-of transit stellar spectra simulating the star with one dominant circular spot as in Eqs. \eqref{model_out} and \eqref{spotted_profile} with different combinations of $d$, $ff$ and $T_s$. Similarly to Paper I, the term “spectrum” refers to both photometric and spectral measurements. In the 3-d parameters space $d-ff-T_s$, the radial distance $d$ of the spot spans the interval $[0\,R_* - 0.95\,R_*]$ with a step of $\delta d = 0.05\,R_*$, the filling factor $ff$ of the spot is defined in the interval $[0.001 - 0.5]$ with a variable step ($\delta ff=10^{-k}$ in the range $ff \in [10^{-k}-10^{-k+1}]$ with $k$ integer going from 3 to 1), while the spot temperature $T_s$ spans the range $[4000\,\si{\kelvin} - 6000\,\si{\kelvin}]$ for HD 17156, $[3000\,\si{\kelvin}-4700\,\si{\kelvin}]$ for HAT-P-11 and $[3000\,\si{\kelvin}-4200\,\si{\kelvin}]$ for K2-21 with a step of $\delta T= 100$ \si{\kelvin}. Tab. \ref{grid_out} shows a summary of the spot parameters used to build the 3-d grids. The lower limit for the spot temperature ($T_s=3000$ \si{\kelvin}) is imposed according to the grid of models \texttt{PHOENIX\textsubscript{-}2012\textsubscript{-}13} implemented in ExoTETHyS. A circular spot having filling factor $ff$, has an elliptical projection on the star disk whose eccentricity depends on its distance $d$ from the star disk center (see Appendix \ref{spot_fraction}), so its effective filling factor, i.e. projected on the stellar disk, will be $ff_{eff}=ff\sqrt{1-d^2}$. Moreover, we emphasize that spots in our grids with $ff<\frac{1-d}{1+d}$ are entirely on the stellar disk, the others partially. Each grid is simulated at ARIEL Tier 2 spectral resolutions with the ArielRad simulator (see Section \ref{ArielRad}) and each spectrum in the grid is in unit of counts (electrons) per second. \\
 
    \begin{table*}
        \centering
        \caption{Summary of the spot parameters used to build the 3-d grid of out-of transit stellar spectra for the three reference stars. The metallicity and the surface gravity for both the spot and the quiet photosphere are set to the solar ones. $T_*$ is taken from the catalogue of \protect\cite{Billy}.}
        \begin{adjustbox}{width=0.7\textwidth}
        \begin{tabular}{c|ccccc}
        \hline
        \hline
             Star     & Spectral Type &   $T_*$   &   $T_s$      &     $ff$       &        $d$   \\
                      &               &   (\si{\kelvin})   &   (\si{\kelvin})      &     (\%)       &     (R$_*$)  \\
        \hline
        HD 17156     &       G0      &  $6040$   &  $4000-6000$  &   $0.1-50$  &      $0-0.95$\\
        HAT-P-11     &       K4      &  $4780$   &  $3000-4700$  &   $0.1-50$  &      $0-0.95$\\
        K2-21        &       M0      &  $4222$   &  $3000-4200$  &   $0.1-50$  &      $0-0.95$\\
         \hline
        \end{tabular}
        \end{adjustbox}
        \label{grid_out}
    \end{table*}
    We use these grids of models to fit a "noisy" out-of-transit spectrum and to derive the best-fit spot parameters that minimize the residuals between the observed spectrum and our grid of models, similarly to what was done in Paper I. The fitting algorithm is implemented with the \texttt{lmfit} python package (\citealt{lmfit}), that provides simple tools for non-linear optimization problems. The fitting algorithm uses as input model a function that interpolates the spectra in the grids, it computes the residual $S$ between the observation $F_\lambda^{obs}$ and the interpolated model $M_\lambda (d,ff,T_{spot})$ as follows:
    \begin{equation}
        \label{residuals}
        S=\sum_{\lambda}(F_\lambda^{obs}-M_\lambda (d,ff,T_{spot}))^2
    \end{equation}
    and minimizes it with the least-squares method in order to find the best-fit values of the spot parameters.
    We use a quadratic interpolation, implemented with the \texttt{scipy} python package (\citealt{scipy}). The interpolation of the grids makes the solution for the best-fit parameters continuous, therefore it is not constrained on the grid nodes. We impose boundary conditions for the spot parameters to be fitted, i.e. that the best-fit solution is within our 3-d domain (the boundaries are reported in Tab. \ref{grid_out}).\\
    
    Our method is tested on realistic ARIEL observations, simulated from a stellar spectrum with an assigned combination of the spot parameters ($d_{input}$, $ff_{input}$, $\Delta T_{input}=T_*-T_S$), where the noise level is generated with the Eq. (5) of Paper I, by assuming $t_{exp}$ = 100 \si{\second}. For each combination of values for $d_{input}$, $ff_{input}$, $\Delta T_{input}$, we simulate $1\,000$ noisy spectra and for each spectrum we run our fitting algorithm by taking care to normalize the simulated spectra (in unit of counts) and the spectra in the grids (in counts/sec) to their integrated fluxes. The method is tested on different combinations of the input spot parameters: $d_{input}=[0, 0.2, 0.4, 0.6, 0.8]\,R_*$, $ff_{input}=[0.01,0.02,0.05,0.1]$, $\Delta T_{input}= [500, 1000, 1500]\,K$\footnote{For K2-21 the case with $\Delta T_{input}=1500\,K$ cannot be tested due to the lower limits of the \texttt{PHOENIX\textsubscript{-}2012\textsubscript{-}13} grid.}. \\ 
    For all the explored cases, we confirm the anti-correlation between the estimated spot filling factor and its temperature contrast found in Paper I; in addition, the derived filling factor is correlated with its estimated distance from the center, because the same result is obtained by a spot closer to the limb with a larger filling factor.\\
    
\begin{figure*}
        \centering
        \includegraphics[width=\textwidth]{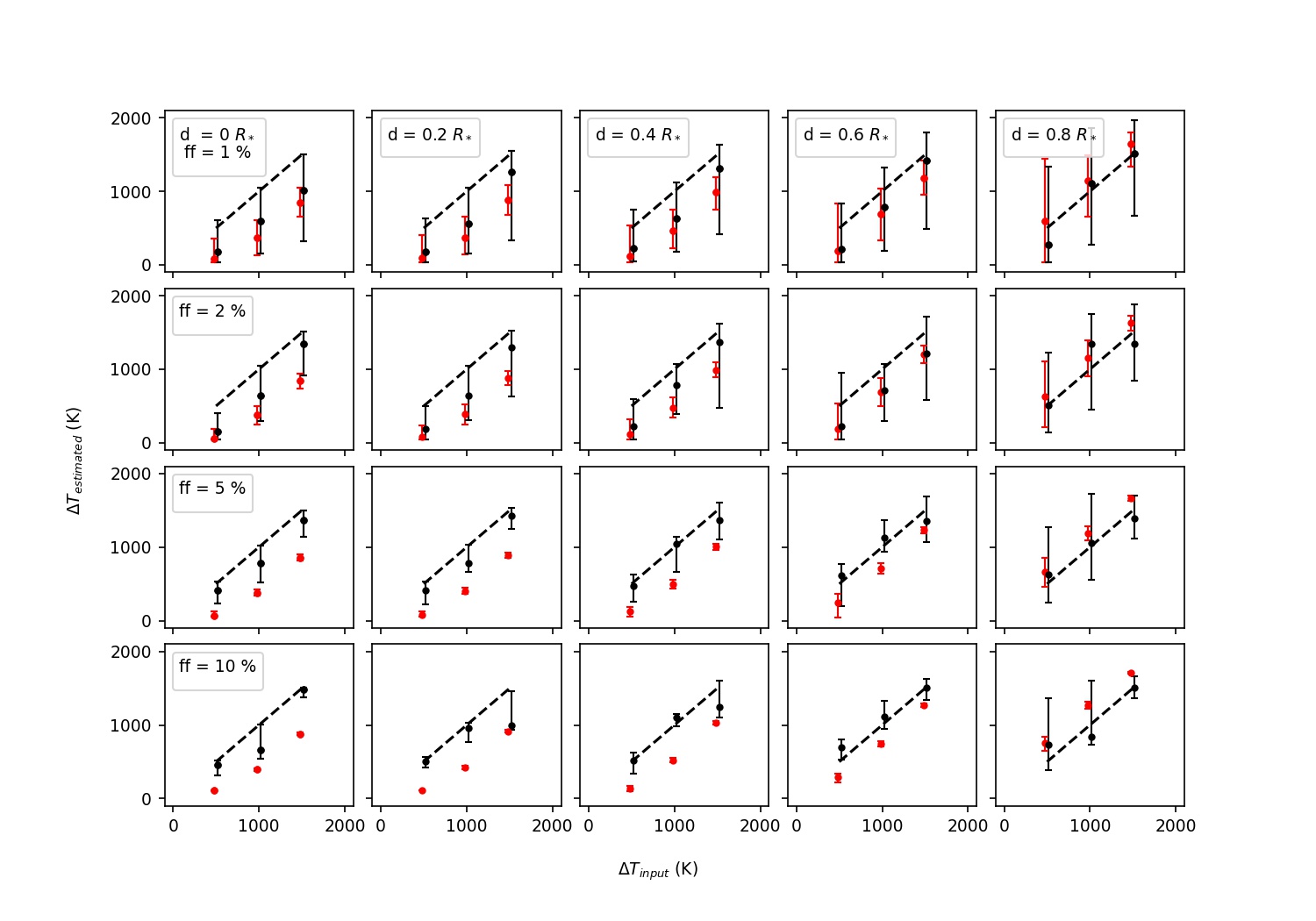}
        \caption{Spot temperature contrasts estimated from our fitting algorithm vs. input spot temperature contrasts for the star HD 17156, when testing cases with different filling factors and spot impact parameters. For each spot, $1\,000$ random realizations of noisy spectra with 100 \si{\second} exposures, are simulated. Plots in the same column are obtained from spots at the same distance from the center (specified in the upper legends), plots in the same row are obtained from the same input $ff$ (shown in the left legends). The black points and the black bars show the median values, the 16\textsuperscript{th} and 84\textsuperscript{th} quantiles of the best fit temperatures when fitting the $1\,000$ random realizations for the spot impact parameter, filling factor and temperature. The red points and the red bars refer to the best fit spot temperatures when fitting the resulting spectra by neglecting the limb darkening. The black and red  bars are horizontally shifted for clarity. The black dashed line shows the expected temperature contrasts in the absence of noise.}
        \label{fig_example}
    \end{figure*}

    As an example, Fig. \ref{fig_example} shows the statistical analysis of the results of our fitting algorithm, applied to the simulations described above for the star HD 17156: the black points and the black bars show the median values, the 16\textsuperscript{th} and 84\textsuperscript{th} quantiles of the $1\,000$ best fit temperatures, respectively. The spot temperature distribution is peaked around the input spot temperature when the spot filling factor is greater than 5\% and its distance from the star center is greater than 0.4 $R_*$. In the other cases, the spot temperature contrasts are overestimated or underestimated compared to the input (black dashed line), with a strong anti-correlation with the estimated filling factors. As expected, the algorithm gives reliable results when the inhomogeneities are strong as in the cases of high contrast/high filling factor spots or close to the limb. \\ 
    
    In order to understand which are the intervals of the parameters where it is necessary to take into account the limb darkening, we perform a fit of each spectrum, simulated with a realistic limb darkening profile, with a 2-d grid of spectra built with spots of different temperatures and filling factors, and limb darkening effect neglected. In order to compare the results with those obtained from the 3-d fitting procedure, we should stress that the 2-d fitting algorithm gives as output an effective filling factor, already projected on the stellar disk, while the 3-d fit returns as best-fit filling factor the ratio between the circular spotted area and the projected star disk. To compare the two cases we have to scale the best-fit filling factor from the 3-d fitting procedure for $\sqrt{1-d_{input}^2}$, being $d_{input}$ the distance of the input spot from the star center. The results are reported in Fig. \ref{fig_example}, where the red points and red bars represent the median values, the 16\textsuperscript{th} and 84\textsuperscript{th} quantiles of the of the $1\,000$ retrieved temperatures from the 2-d fitting algorithm. A comparison between the results from the two algorithm shows that when the input spot has a $ff<0.02$ and is at $d = 0.6\,R_*$, the median values of the distributions obtained from the two fits tend to coincide, thus showing that $d = 0.6\,R_*$ is an "effective" distance in which the two methods give similar results, i.e. the limb darkening effect is negligible. For larger filling factors, the fitting performed by neglecting the limb darkening effect, introduces systematic offset in the recovered values. Analogous results are obtained for the other two stars in our sample.
    
    In Section \ref{retrieval_planet}, we use this fitting algorithm to model the stellar activity due to the presence of spots from the out-of transit observations in order to extract the planetary transmission spectrum. 

\subsection{Method to retrieve the planet's transmission spectrum and the impact parameter}\label{retrieval_planet}
    As done in Paper I, we develop a method to extract the transmission spectrum of a transiting planet in the presence of star spots and we test it on ARIEL simulations. Since ARIEL will observe a sample of known extrasolar planets, we can work under the reasonable assumption that the orbital period $P$ and the transit mid time are known, so they can be considered fixed in the fitting process. Furthermore, we are assuming circular orbits, and if we have an accurate estimate of the stellar mass, we can derive the semi-major axis $a$ from the Kepler's third law and we can also assume it as fixed. From the photometry of the planetary transits, we can also have some constraints on the impact parameters and the radii of the planets that ARIEL will observe, but in our method we consider them free parameters as we want to assess if the presence of the spot may have a significant effect on the determination of these two parameters and if our method can remove these systematics.\\
    Our method is based on the hypothesis that the spotted area can be assumed constant during both the transit and the immediate out-of transit and this is a reasonable hypothesis for the majority of systems as the spot variability is modulated on time scales much longer than that of the typical transit. Under this hypothesis, if the planet transits out of the spot, the latter can be removed when subtracting the out-of transit stellar flux $F_\lambda^{out}$ from the transit observation $F_\lambda ^{in}$. The difference $F_\lambda^{out}-F_\lambda ^{in}$ gives the unperturbed photospheric flux blocked by the transiting planet at each wavelength. Our method consists in comparing $F_\lambda^{out}-F_\lambda ^{in}$ with a grid of models each describing the variation of the photospheric flux $\delta_\lambda(t)\cdot P_\lambda (T_*)$ over time due to a transiting planet with an assigned planetary radius $R_P$ and an assigned impact parameter $b$. 
    The terms $\delta_\lambda(t)\cdot P_\lambda (T_*)$ are computed with the \texttt{pylightcurve} package (see Section \ref{simulate_light_curves}) by varying the position of the planet every $100$ \si{\second}, i.e. the same step used to simulate the transits.
    We emphasize that the grid is strongly dependent on the adopted set of stellar parameters ($T_*$, $\log{g}$, $[Fe/H]$) used to model the quiet photosphere of the target to be observed, so the method requires an accurate characterization of the stellar parameters. On the other side, the advantage of this approach is that we do not consider any influence of the spot on the limb darkening coefficients because we are just modeling the non-spotted flux.\\
    In our grids, $b$ spans the interval $[0-1]$ with a step $0.1$ for $b<=0.7$ and $0.01$ for $b>0.7$: the grids are finer for high impact parameters as the intensity decrease due to the limb darkening is very steep in this region.
    The planetary atmospheres simulated here extend for about 10 scale heights, so the grids span an interval of $[0-15]$ scale heights with steps of $1.5$ scale heights above the photometric estimate of the planetary radius.
    
    We use the grid, integrated over the entire ARIEL wavelength coverage, to fit the white time series $F_\lambda^{out}-F_\lambda ^{in}$ and to derive the best-fit effective planetary radius and the planet impact parameter $b$, by using the same algorithm described in Section \ref{retrieval_params}. Then, we fix the value of the planet impact parameter $b$ previously found, by linearly interpolating the grid, and we perform a wavelength dependent fitting of $F_\lambda^{out}-F_\lambda ^{in}$ to derive the best-fit planetary radius at each wavelength (i.e. the planet transmission spectrum).
    When the planet crosses a spot, it will block a fraction of the unperturbed photospheric flux and a fraction of flux from the spot ( \citealt{Espinoza,Sanchis_Ojeda}), thus producing an upward bump in the transit light curve. At first we establish a criterion to detect and remove the bump from the wavelength integrated light curve (described in Section \ref{spot_crossings}). We perform a fit of the white time series $F_\lambda^{out}-F_\lambda^{in}$, where the bump is removed, to derive the planet impact parameter with the method described above. Then, we perform a chromatic fit of $F_\lambda^{out}-F_\lambda^{in}$ to derive the planet transmission spectrum. In the last fitting the bump is included, but, for the cases of spot crossing events, we further simplify our approach, by assuming that we have no emission from the spot, so that: in the transit data points out-of the spot crossing, the planet will block a fraction of flux from the photosphere equal to $\delta_\lambda(t)\cdot P_\lambda (T_*)$, while during the spot crossing only a portion will be occulted. We indicate the fraction of the missing photospheric flux as $g_\lambda (t)$ and we model this wavelength-dependent factor with a Gaussian profile:

    \begin{equation}
        g_\lambda (t)= A\cdot\exp{\left(\frac{(t-t_0)^2}{B^2}\right)}
    \end{equation}
    where the three free parameters $A$, $t_0$ and $B$ are wavelength-dependent. $A$ represents the maximum fraction of unperturbed photospheric flux non blocked by the planet, $t_0$ is the time of the maximum spot occultation, while $B$ is the time length of the spot occultation. The factor $g_\lambda(t)$ is expected to depend on wavelength. In this case from the fit we obtain 4 parameters for each wavelength ($R_P$, $A$, $t_0$ and $B$). In Section \ref{results}, we show the results of our fitting algorithm applied to the three reference planetary systems.
    
    The method described above is suggested to correctly derive the radius and the impact parameter of a planet transiting in front of a spotted star. In order to quantify the impact of the spots on the determination of these two parameters, we fit the transit observation $F_\lambda^{in}$ (normalized to its out-of-transit) by ignoring the presence of the spots and by considering only the radius and the impact parameter of the planet as free parameters in the fitting. The specific cases are discussed in Section \ref{results}.
    
\section{Results}\label{results}
    We apply our method to planetary transits in which the star is simulated with spots of different sizes, temperatures and positions on the stellar disk, defined by their latitudes and longitudes. For each simulated transit light curve we consider the integrated out-of-transit exposures, i.e. the combined out-of transit spectra, and we fit the resulting spectrum as described in Section \ref{retrieval_params} in order to estimate the spot parameters, and we apply the method presented in Section \ref{retrieval_planet} to retrieve the planetary transmission spectrum and its impact parameter, for both crossed and non-crossed spots. For each test, we simulate $n$ transits observations (reported in Tab.\ref{parameters}), being $n$ the number of transits, computed by \cite{Billy}, needed to observe a primordial atmosphere in each target at Tier 2 resolution, with $S/N=7$. The same choice has been done in Paper I.

    \subsection{Transits out of the spots}\label{transit_out_spot}
    For each transiting system, we simulate different transit observations with the star having one dominant spot and for two different values of temperature contrast $\Delta T_{input}=500,\, 1000\,K$ and two different values of filling factor $ff_{input}=0.02,\,0.05$. The three simulated planets have an orbital inclination $i<90$\si{\degree} (see Tab. \ref{parameters}), therefore pylightcurve simulates their transits in the southern hemisphere of the stars; since we are analyzing the case of transits out of the spots, 
    we assume that  the spot is on the northern hemisphere, at a latitude $\theta=20$ \si{\degree}. We assume an initial longitude of the spot $\psi=-20$\si{\degree} and for each combination $\Delta T_{input}$, $ff_{input}$ we randomly extract $n$ transits from $1\,000$ consecutive transits, by assuming that the spot longitude evolves with the stellar rotation and that the spot is on the visible hemisphere of the star. This procedure allows to sample different distances of the spot from the star center in each transit. The stellar rotation axis is assumed perpendicular to the line of sight and the stellar rotation periods for our reference stars are shown in Tab. \ref{parameters}.\footnote{For K2-21, we do not have a clear evidence of flux modulation due to spots (\citealt{rotation_K2_21}), however we assume for this target $P_s=20$\si{\day}.} In each simulated transit the spot filling factor is randomly extracted from Gaussian distributions of $ff$, peaked at $\mu_{ff}=0.02,\,0.05$ and standard deviation $\sigma_{ff}=0.01$. For negative values of $ff$, the extraction is iterated until all the extracted values are positive.
    It is easy to show that the distance $d$ of the spot can be derived from its latitude $\psi$ and its longitude $\phi$ from the relation:
    \begin{equation}
        \label{spot_impact_param}
        d=\sqrt{1-\cos^2{\phi}\cos^2{\psi}}
    \end{equation}
    The impact parameter $b$ of each transiting planet is shown in Tab. \ref{parameters} and is derived as:
    \begin{equation}
        \label{impact_parameter}
        b=\frac{a}{R_*}\cos{i}
    \end{equation}
    where $a$ and $i$ are the orbital semi-major axis and the orbital inclination, respectively. We use the integrated out-of transit observations to derive the spot properties as described in Section \ref{retrieval_params}. Tab. \ref{out_transit} shows the best-fit spot parameters retrieved in some specific cases. We integrate the out-of-transit spectrum to increase the S/N ratio and to reduce the uncertainty on the estimated spot parameters. The errors on the parameters in Tab. \ref{out_transit} are the formal errors derived from the fitting algorithm, which are probably underestimated and not include residual systematic effects.
    \begin{table*}
        \centering
        \caption{Summary of the best-fit (bf) spot parameters retrieved from specific out-of transit observations, simulated with different input spots. The errors associated to the best-fit values come from the fit. The values of $d_{input}$ have been obtained with Eq.\eqref{spot_impact_param}.}
        \begin{adjustbox}{width=0.7\textwidth}
        \begin{tabular}{c|cccccc}
        \hline
        \hline
             Star     & $d_{input}$ &   $ff_{input}$       &  $\Delta T_{input}$   & $d_{bf}$ &   $ff_{bf}$       &  $\Delta T_{bf}$   \\
                      &   ($R_*$)   &      (\%)            &   (\si{\kelvin})      & ($R_*$)  &     (\%)    &    (\si{\kelvin}) \\
        \hline
        
        \multirow{2}{*}{HD 17156} & 0.39 & 2.0 & 500 &  $0.35\pm 0.03$ & $1.72\pm 0.08$ & $573.5\pm 21.0$\\
        & 0.61 & 1.7 & 1000 & $0.71\pm0.02$ & $2.46\pm0.25$ & $783.1\pm51.5$\\
        \cline{2-7}
        \multirow{2}{*}{HAT-P-11} & 0.38 & 2.7 & 500 & $0.36\pm0.01$ & $2.63\pm0.02$ & $500.7\pm1.8$\\
        & 0.76 & 2.5 & 1000 & $0.77\pm 0.01$ & $2.60\pm 0.05$ & $990.5\pm 3.7$\\
        \cline{2-7}
        \multirow{2}{*}{K2-21}  &  0.61 & 1.3 &  500 &  $0.66\pm 0.03$ & $1.54\pm 0.14$ & $413.5\pm 22.2$\\
        &  0.36 & 3.5 & 1000  &  $0.39\pm 0.01$ & $3.60\pm 0.02$ & $995.2\pm 3.8$\\
         \hline
        \end{tabular}
        \end{adjustbox}
        \label{out_transit}
    \end{table*}

\begin{figure}
        \centering
        \includegraphics[width=\columnwidth]{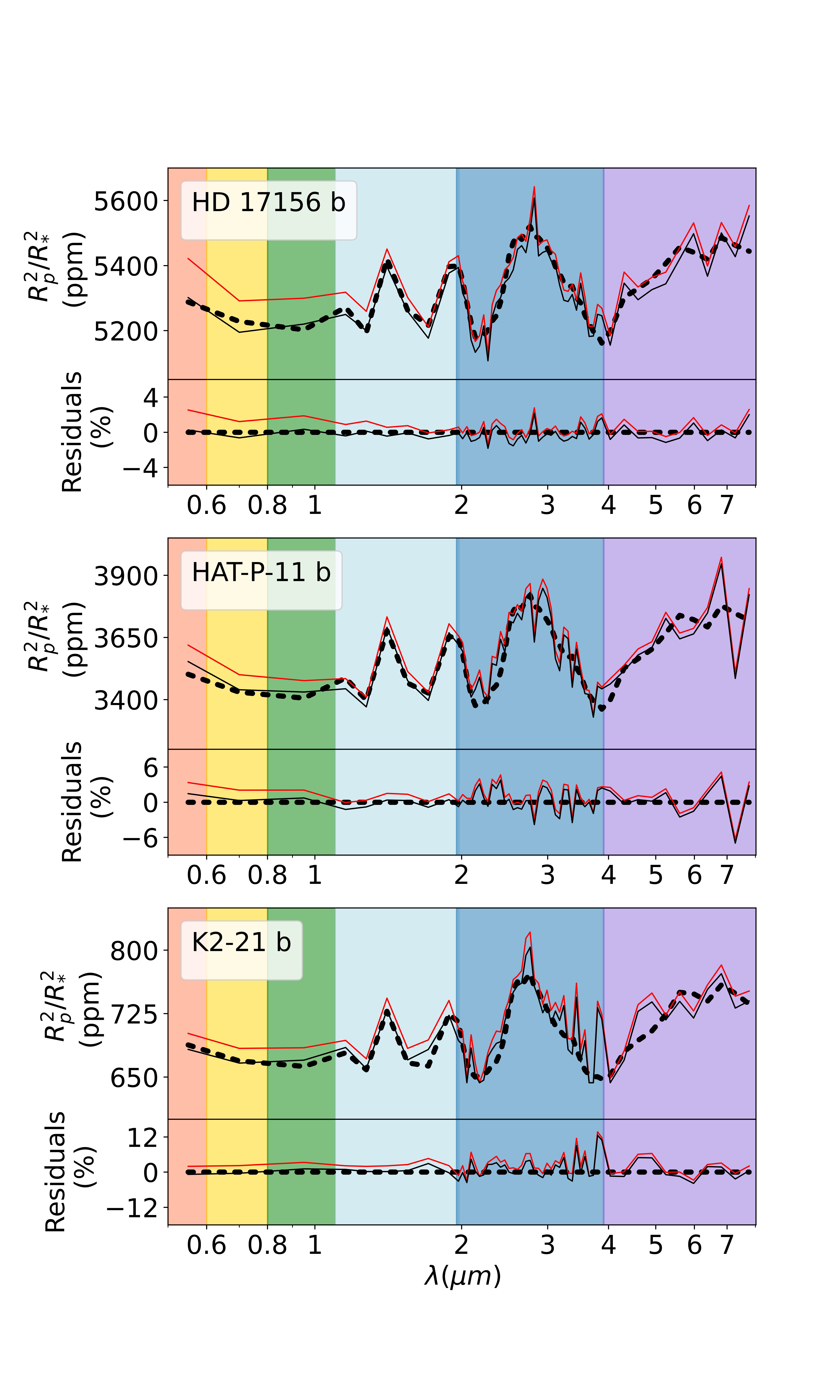}
        \caption{Retrieved planetary spectra, uncorrected (red lines) and corrected (black continuous lines) for the presence of the unocculted spot, averaged over $n$ transits ($n$ is reported in Tab.\ref{parameters}). In each simulated observation, the spot temperature contrast is $\Delta T=1000\,K$, while the filling factor is randomly extracted from normal distributions with $\mu_{ff} = 5\%$  and $\sigma_{ff} = 1\%$. The dashed lines are the input planetary transmission spectra. The spectra are at Tier 2 resolutions. For each case the residuals are expressed as percentage deviations from the input transmission spectrum. The correction for the unocculted spot is stronger below 2\si{\micro\meter}.}
        \label{retrieved_pl_spectra}
    \end{figure}

    At first, we quantify the influence of the spot on the extraction of the planetary transmission spectrum and of its impact parameter. We find that for $\Delta T_{input}=500\,K$, or lower, the retrieved transmission spectrum of the transiting planet is not contaminated by the presence of the spot due to the noise level; on the contrary, for $\Delta T_{input}=1000\,K$, or higher, we find an overestimate of the planetary radius at each wavelength of a few percents, with a chromatic dependence. In particular, the effects are stronger below 2\si{\micro\meter}, where the spot-photosphere contrast increases, and negligible in the near IR. These results are shown in the three panels of Fig. \ref{retrieved_pl_spectra}, where the input planetary transmission spectrum is represented with a dashed line and the retrieved uncorrected spectrum with a continuous red line. The latter is the mean of the $n$ extracted spectra for the brightest targets (HD 17156b and HAT-P-11 b), the median of the $n$ extracted spectra for the faintest target (K2-21 b). 
    In all the cases the planetary radius is overestimated above the errors: the planetary radius of HD 17156 b is overestimated in the three photometric channels by about 100 ppm to be compared with a measurement error of about 22 ppm; for HAT-P-11 b the overestimate is by about 85 ppm vs. an expected error of about 20 ppm; for K2-21 b the overestimate is by about 35 ppm vs. an expected error of about 4 ppm. If we apply the method described in Section \ref{retrieval_planet}, we can recover the planetary spectrum within the noise level and remove the systematic overestimate (see black continuous lines in Fig. \ref{retrieved_pl_spectra}).
    
    Fig. \ref{retrieved_pl_spectra} clearly shows that a correction of the planetary spectrum is necessary for the case with $ff=0.05$ and $\Delta T=1000\,K$ because ignoring the presence of the spot can leave residuals of a few percent below $2$ \si{\micro\meter} that can mimic the typical slopes of Rayleigh scattering in the planetary transmission spectrum (see, e.g., \citealt{Pont_2013, Sing_2011}).\\

    Moreover, for all the explored cases we find that the spot presence do not significantly affect the retrieval of the planetary impact parameter; in particular, we find that the planetary impact parameter is retrieved with a precision level below 1\% for both HD 17156 b and HAT-P-11 b, and of a few percents for K2-21 b whether we correct or not for the presence of the spots, thus suggesting that the spots do not significantly affect the derivation of the planet impact parameter. Probably, this is a consequence of fixing all the other orbital parameters of the transiting system (\citealt{Ballerini}).\\
    
    We have verified that the capability of the method to correct the planetary spectrum for the presence of the spot is not dependent on its geometrical and physical properties, or on the number of simulated spots (see Section \ref{robustness_of_the_model}), as the presence of the spot is removed through the difference $F_\lambda^{out}-F_\lambda^{in}$, and it depends only on our capability to model the unperturbed photospheric flux.\\
    
    In Appendix \ref{Discussion}, we will show the application of the method presented in Paper I to extract the planet transmission spectrum from the light curves simulated in the present work, so as to estimate the bias introduced by neglecting the limb darkening in the retrieval of the transmission spectrum.
    
    \subsection{Robustness of the method with respect to the spot distribution}\label{robustness_of_the_model}
    
    Our model of stellar activity is based on the assumption that the visible hemisphere of the star is dominated by the presence of a main spot with a given temperature and filling factor, ignoring that the star may have different spots at different position on its disk. In order to understand if and how such an assumption affects our results, in the following we simulate a star with two spots having different sizes and at different positions on the stellar disk and we apply our method based on one spot to recover the spot "effective" parameters from the out-of transit and the planetary transmission spectrum.\\
    
    We simulate transit observations where the two input spots have $\Delta T_{input} = 1000\,K$, are at latitudes $\theta_1=30$\si{\degree}, $\theta_2=-10$\si{\degree} and initial longitudes $\psi_1=-20$\si{\degree} and $\psi_1=10$\si{\degree}. For each transiting system, we simulate $n$ observations (see Tab.\ref{parameters}), where the filling factors of the two spots are randomly extracted from normal distributions with mean $\mu_{ff}=2\%$ and standard deviation $\sigma_{ff}=1\%$. As before, the $n$ transits are randomly extracted from $1\,000$ consecutive observations, by assuming that the two spots co-rotate with the stellar surface and that no differential rotation is simulated.\\
    
    The spot parameters are derived by comparing the integrated (sum of the) out-of-transit observations, simulated with two spots, with our grid of out-of-transit spectra, built by assuming the presence of one dominant  spot on the star disk, therefore the estimated parameters are "effective" parameters. Tab.\ref{out_transit_2_non_crossed_spots} shows a brief summary of the retrieved spot parameters in some specific transits. We find that the method can: confidently recover the input spot temperature contrast, retrieve also an effective impact parameter that mediates the distances of the two input spots, and derive a best-fit effective filling factor roughly given by the sum of the two input filling factors.
    
\begin{figure}
        \centering
        \includegraphics[width=\columnwidth]{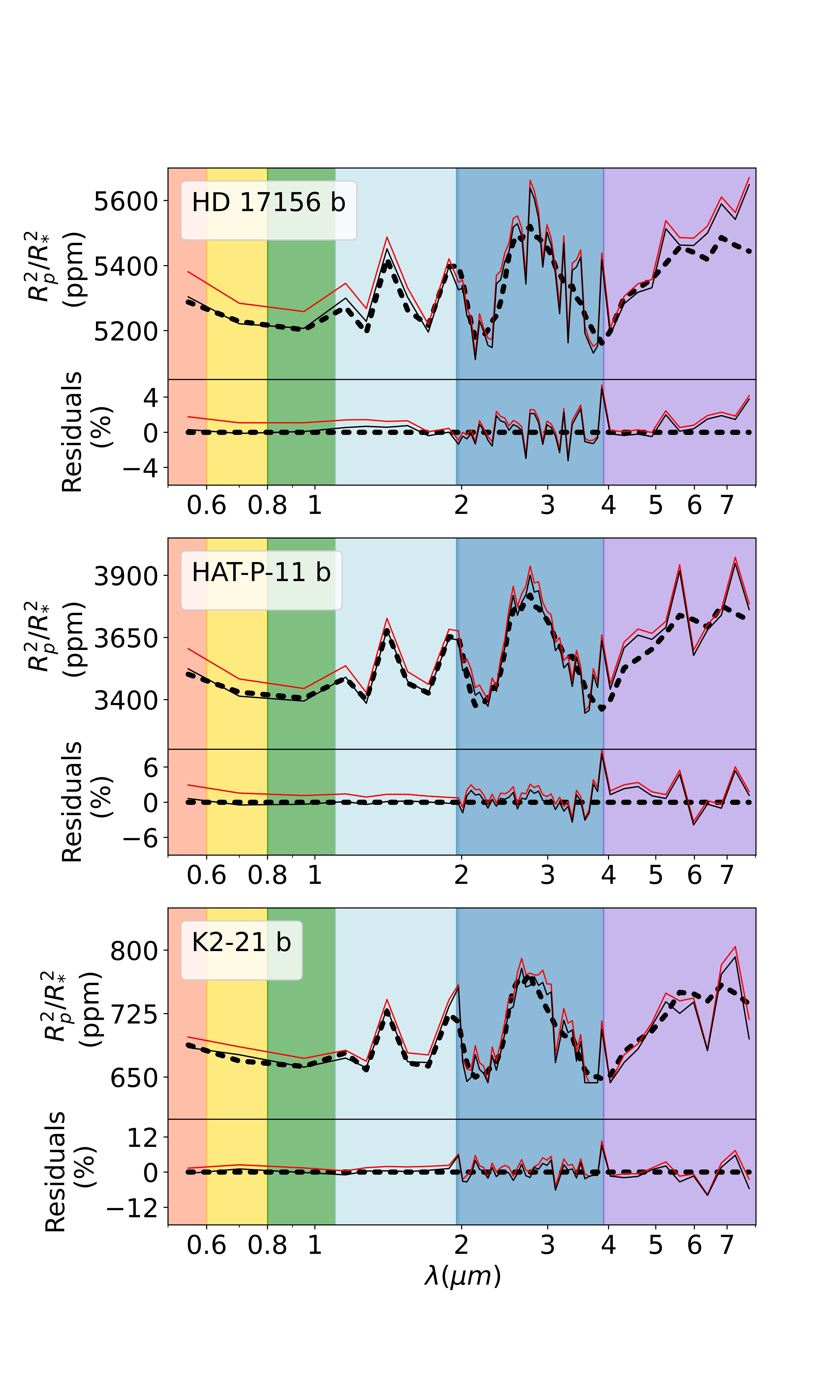}
        \caption{As in Fig.\ref{retrieved_pl_spectra} for planetary transits in which the star is simulated with two uncrossed spots having $\Delta T= 1000\,K$, $ff$ randomly extracted from Gaussian distributions peaked at $\mu_{ff}=2\%$ and with $\sigma_{ff}=1\%$ and random positions.}
        \label{retrieved_pl_spectra_two_unocculted_spots}
    \end{figure}

    \begin{table*}
        \centering
        \caption{Summary of the best-fit (bf) spot parameters retrieved from specific out-of transit observations when the star is simulated with two spots (1,2) and its spectrum fitted with a one-dominant-spot model. The errors associated to the best-fit values come from the fit.}
        \begin{adjustbox}{width=0.9\textwidth}
        \begin{tabular}{c|cccccccc}
        \hline
        \hline
             Star     & $d_{input,1}$ &   $d_{input,2}$ & $ff_{input,1}$ &   $ff_{input,2}$       &  $\Delta T_{input}$   & $d_{bf}$ &   $ff_{bf}$       &  $\Delta T_{bf}$   \\
                      &   ($R_*$)   &      (\%)            &   (\si{\kelvin})      & ($R_*$)  &     (\%)    &    (\si{\kelvin}) \\
        \hline
        
        HD 17156 & 0.85 & 0.47 & 2.7 & 2.0 &  1000 & $0.73\pm 0.01$ & $5.07\pm 0.12$ & $913.9\pm 14.6$\\

        HAT-P-11 & 0.84 & 0.45 & 0.8 & 1.6 & 1000 & $0.56\pm 0.01$ & $2.19\pm 0.04$ & $998.5\pm 4.0$\\

        K2-21 & 0.54 & 0.84 & 2.3 & 1.0& 1000 & $0.62\pm 0.01$ & $3.11\pm 0.03$ & $979.5\pm 4.8$\\
         \hline
        \end{tabular}
        \end{adjustbox}
        \label{out_transit_2_non_crossed_spots}
    \end{table*}
    Fig.\ref{retrieved_pl_spectra_two_unocculted_spots} shows the retrieved planet spectra for the three planets simulated here, corrected (black continuous line) and uncorrected (red line) for the presence of the two spots. Also for these cases, the method can extract the planetary transmission spectrum within the noise level. We have found similar results also for more than two spots in the star disk, thus suggesting that the method can reliably extract the planet's atmospheric modulation even for a more realistic spots' distribution. We confirm that, also in these cases, the spots do not influence the recovered planet impact parameter.
    
    \subsection{Spot crossing events}\label{spot_crossings}
\begin{figure}
        \centering
        \includegraphics[width=\columnwidth]{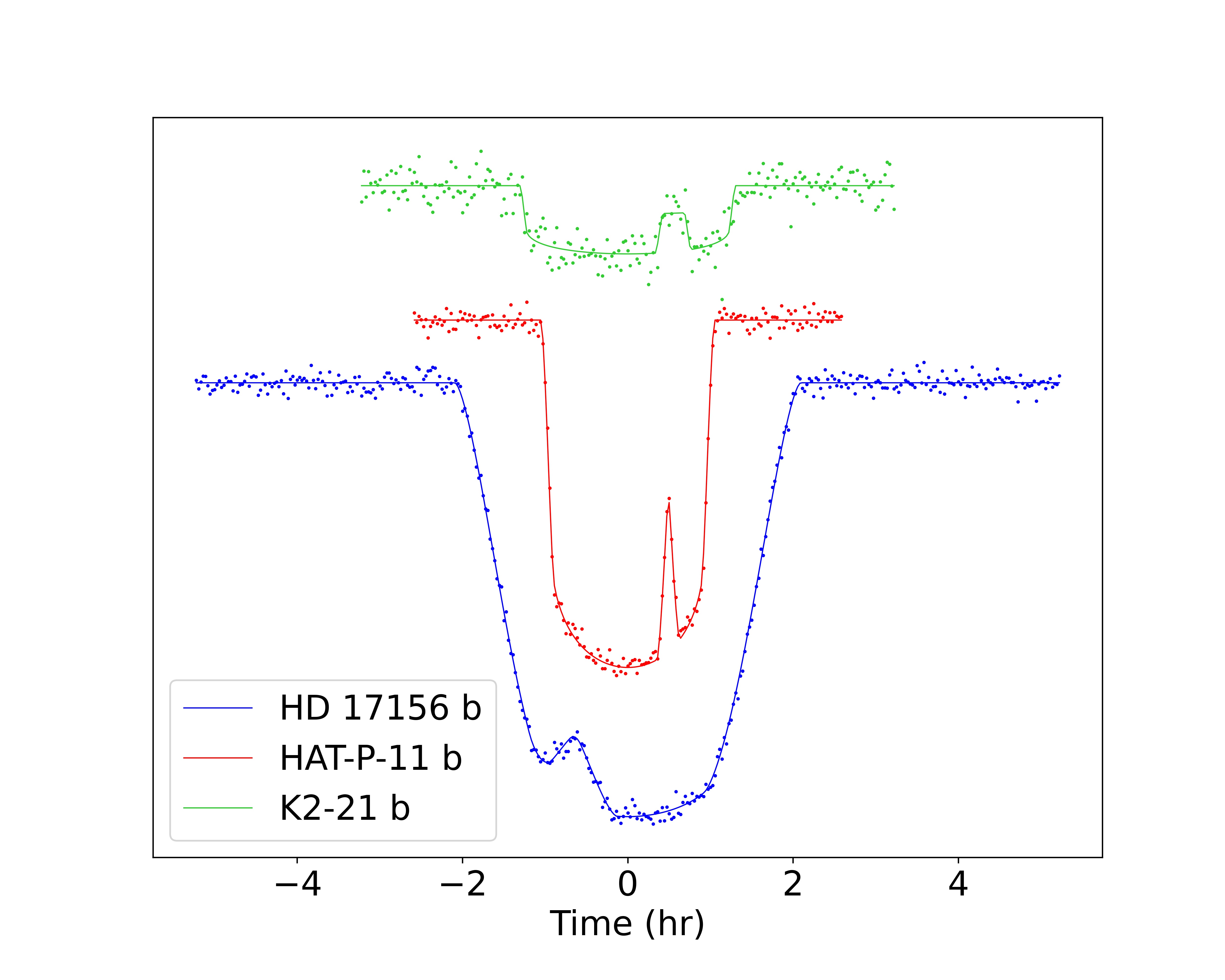}
        \caption{ Synthetic transit light curves with spot crossings for the three planetary systems specified in the legend. The light curves are integrated over the entire ARIEL wavelength coverage. The crossed spots have $\Delta T_{input}=1\,000\,K$ and size comparable to $ff\sim1\%$.}
        \label{lc_with_bump}
    \end{figure}

    When a planet crosses a large and cool enough spot and the star is very bright, it is possible to detect a maximum during the transit (bump). 
    In the following, we generate $n$ transit observations ($n$ is reported in Tab.\ref{parameters}) for each combination star+planet, where the star is simulated with two different spots co-rotating with the stellar surface. The $n$ transits are randomly extracted from $1\,000$ consecutive transits (see Section \ref{transit_out_spot}), and one of the two spots is on the planet transit chord, so that it is crossed by the transiting planet. We set the latitude of the crossed spots (assumed in the southern hemisphere, where the transit takes place) to: $\theta=-65$\si{\degree} for HD 17156 b, $\theta=-30$\si{\degree} for the other two simulated planets. As before, the spot filling factors are randomly extracted from Gaussian distributions peaked at $\mu_{ff}=1\%$ and $\sigma_{ff}=1\%$.
    As an example, Fig. \ref{lc_with_bump} shows three simulated transit light curves for our reference planets with spot crossing events.
    
\begin{figure}
        \centering
        \includegraphics[width=\columnwidth]{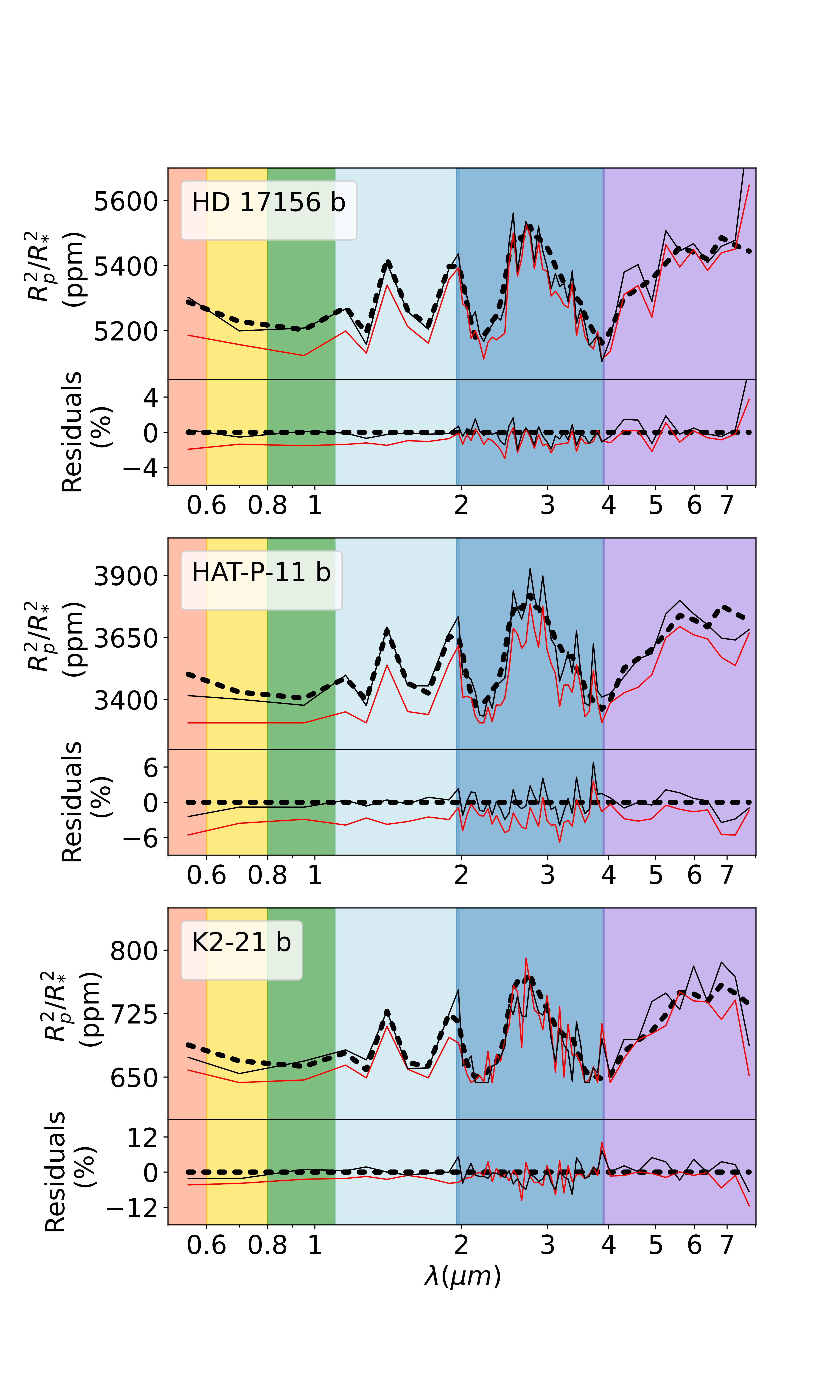}
        \caption{As in Fig.\ref{retrieved_pl_spectra_two_unocculted_spots}, where one of the two simulated spots is crossed by the planet.}
        \label{retrieved_pl_spectra_two_occulted_spots}
    \end{figure}

   Since we want to take into account the noise effect on ARIEL observations, we first have to establish a criterion for the detectability of the bump in the simulated transit light curves. To this end, we fit each transit, integrated over the ARIEL wavelength coverage, by ignoring the presence of the spots, to derive the combination of $b$ and $R_P/R_*$ that best match the simulated light curve; then, if we have $m$ consecutive points where the residuals between the observed light-curve and the best-fit model that are $k\cdot\sigma$ beyond the median value of the residuals, being $\sigma$ the standard deviation of the residuals, we establish that the bump is detected and we correct the transit observation for both occulted and unocculted spots as described in Section \ref{retrieval_planet}. In all the other cases, we establish that the bump is undetected and we correct only the unocculted spots. The parameters $m$ and $k$ are strongly dependent on the S/N ratio of the observed targets and the temperature contrast of the crossed spots: for HD 17156 b and HAT-P-11 b we choose $m=2$ and $k=2$ (reasonable assumption for the detection of crossed spots with $\Delta\, T\gtrsim 500\,K$); for K2-21 b we choose $m=4$ and $k=0.5$ (good assumption for spots with $\Delta\, T\gtrsim 1000\,K$). The choice of this criterion is driven by the comparison between the size of the observational errors and the expected size of the bump. In particular, for the brightest stars the probability to have a spurious bump with $m=2$ and $k=2$ is $5\times10^{-4}$, while for the faintest target the choice of $m=4$ and $k=0.5$ gives a probability of a spurious bump of  $10^{-2}$. Since we simulate 184 transits for the latter target, we expect that about 2, out of the 155 detected bumps, are spurious. 
   We apply the method described in Section \ref{retrieval_planet} to derive the planet transmission spectrum. The results are shown in Fig. \ref{retrieved_pl_spectra_two_occulted_spots}, where the black continuous lines represent the values of the retrieved transmission spectra averaged over the $n$ simulated transits for the case with spots having $\Delta T=1000\,K$. For the target K2-21 b, our method cannot reveal the bumps in the transit light curves for wavelengths longer than 2 \si{\micron}, therefore, in this spectral range, the method correct only for the presence of the unocculted spots. The red lines in Fig.\ref{retrieved_pl_spectra_two_occulted_spots} show the retrieved planet spectra when we fit the transit light curves in each of the ARIEL photometric channels and spectral bins by ignoring the presence of both occulted and unocculted spots. The plot shows that the retrieved planetary radius at each wavelength is underestimated compared to the input, thus confirming that the crossed-spots are responsible for the dominant effect when extracting the planetary transmission spectrum. The distortion effects are stronger at shorter wavelength where the spot-photosphere contrast increases.
    
\section{Summary and discussions}\label{conclusion}
    We have developed a method to recover the planetary transmission spectrum in the presence of stellar activity due to star-spots. The method is an extension of the method already presented in Paper I, where we considered a spot dominated scenario, without taking into account the limb darkening effect. The main assumption of Paper I is uniform emission from the whole star disk and from the spot, so the spot is parameterized only by its filling factor and temperature, and the transit depth is dependent only on the relative planetary size $(R_P/R_*)^2$. In this work, we show the need to apply a realistic limb darkening treatment. Our method shows that  both the distance of the spot from the center of the stellar disk and the impact parameter of the transiting planet are crucial parameters to correctly recover the planetary spectrum, requiring the  need to simulate and analyze realistic transit light curves. \\
    
    As for the retrieval of the spot parameters, the method consists in fitting an out-of transit spectrum over a 3-d grid of spectra, where each element represents the star with one dominant circular spot, with different sizes, temperatures and distances from the stellar disk center. The spots are simulated as parts of the photosphere at a lower temperature, with the same surface gravity and abundance of the quiet photosphere. The spectra of the spots and the photosphere are taken from the BT-Settl models (\citealt{Baraffe}) and each spectrum of the spotted star is degraded at ARIEL Tier 2 resolutions with the software ArielRad (see Section \ref{ArielRad}). We use an algorithm to fit a noisy out-of transit spectrum and to recover the spot physical and geometrical properties of the spot.\\
    
    The transmission spectrum of the atmosphere of the transiting planet and its impact parameter are obtained by modeling the unperturbed photospheric flux occulted by the planet, in the assumption that the spot variability is modulated over a time scale that is much longer than the typical transit duration. Under this hypothesis the spot is removed when subtracting the entire transit observation ($F_\lambda^{in}$) to the out-of transit observation ($F_\lambda^{out}$). 
    In particular we fit the observed $F_\lambda^{out}-F_\lambda^{in}$ over a grid of models each describing the photospheric flux blocked by the planet during the transit for planets having different combinations of radii and impact parameters. These grids of time series can be optimized for each target to be observed, by choosing the proper values of radii and impact parameters.
    
    Our method consists in modeling the flux blocked by the transiting planet and this is possible only if: 1) we have a reliable simulator of the instrument; 2) we have an accurate characterization of the stellar parameters (see Paper I for a more detailed discussion). \\
    
    The method is tested with simulations of the three planetary systems: HD 17156 b, HAT-P-11 b, K2-21 b. The results show that the algorithm reliably recovers the spot parameters especially if we use long exposure times equivalent to the integrated out-of-transit observations ($\sim$ a few hours) because they achieve a good S/N ratio of the observation. In particular, we confirm the anti-correlation between the spot filling factor and temperature contrast, we find a correlation between the spot filling factor and its distance from the center of the star disk. Moreover, we find that for a spot close to the limb the method can better constrain its distance, as a consequence of the stronger limb darkening effect. We also show that our method can be applied to cases with a more realistic spots distribution, i.e. with more than one spot in the visible stellar hemisphere, thus demonstrating that our analysis, if applied to observed out-of-transit spectra, can effectively provide a systematic characterization of the photospheric activity as due to the stellar spots, for stars of different spectral types. \\
    
    For each target, we have simulated transit observations where the star has one or two visible spots with various filling factors, temperatures and positions on the star disk, either crossed or not by the planet, and the planet has a primordial atmosphere. The number of transits simulated for each target is taken from the catalogue of \citealt{Billy}.
    We have shown that if we neglect the limb darkening  when we extract the spectrum of a transiting planet, we introduce a systematic offset both in the retrieved spot parameters and in the recovered planetary spectrum (see Appendix \ref{Discussion}), thus demonstrating the importance of including the limb darkening effect.
    The method presented in this work is able to reliably retrieve the atmospheric transmission spectrum and the impact parameter of the transiting planet removing the systematic effect due to spots, for both the cases of occulted and unocculted spots, by averaging $n$ observations.\\
    
    Our approach has been developed for ARIEL but it may be extended to any other mission aimed at observing the atmospheres of transiting extrasolar planets, provided that an accurate simulator of the science payload performances of the mission is available. This is necessary to build the grids of spectra for each potential target. In future works, we plan to apply this analysis to real data observed with the Wide Field Camera 3 (WFC3) of Hubble Space Telescope in order to identify the targets with stronger evidence of stellar magnetic activity and to retrieve the planets atmospheric transmission spectra.
    
\section*{Acknowledgements}
      The authors acknowledge the support of the ARIEL ASI-INAF agreement n.2021-5-HH.0. The authors thank Enzo Pascale and Lorenzo Mugnai from the University La Sapienza of Rome for providing us the ArielRad software simulator. Giuseppe Morello has received funding from the European Union's Horizon 2020 research and innovation programme under the Marie Skłodowska-Curie grant agreement No. 895525.

\section*{Data Availability}
    Data used in this paper are derived from simulations of publicly available codes: \url{https://phoenix.ens-lyon.fr/Grids/BT-Settl/} (\citealt{Baraffe}), \url{https://github.com/ucl-exoplanets/TauREx3_public} (\citealt{TauReX}),  \url{https://github.com/ucl-exoplanets/pylightcurve} (\citealt{pylightcurve}) and the ArielRad software simulator (\citealt{ArielRad}). The specific simulation can be shared on reasonable request to the corresponding author.

\bibliographystyle{mnras}
\bibliography{file}

\appendix

\section{Computing the fraction of spotted annulus}\label{spot_fraction}

\begin{figure}
    \centering
    {\includegraphics[width=0.48\columnwidth]{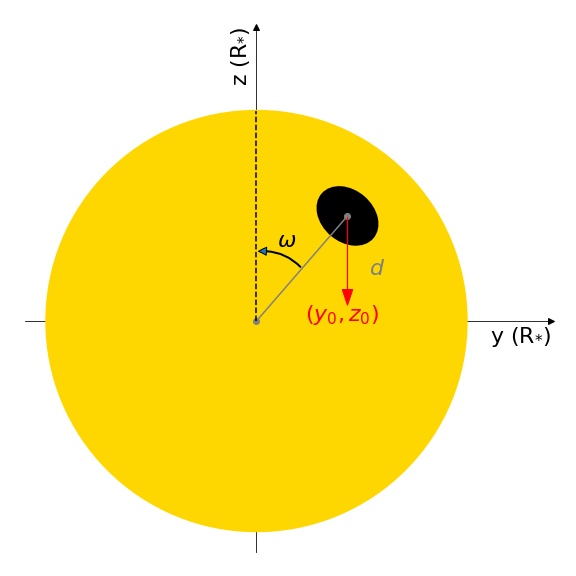}\quad
    \includegraphics[width=0.48\columnwidth]{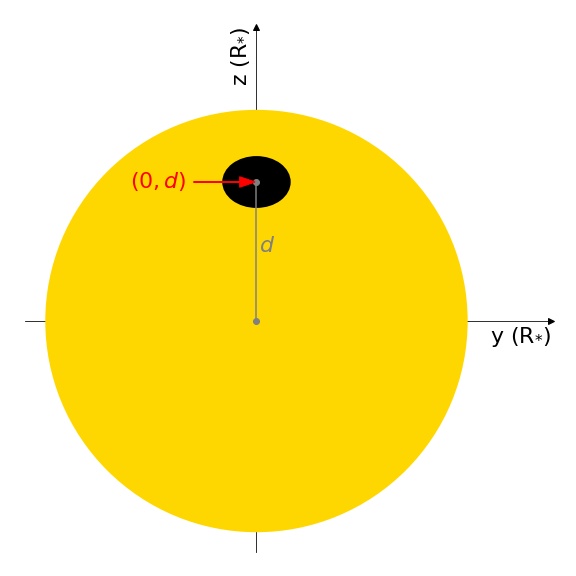}}
    \caption{The left-hand panel shows the spot with its center is in the position $(y_0, z_0)$ of the projected stellar disk. The $x$ axis is outgoing and the observer is at $x=+\infty$. The rotation of the system “star+spot” of an angle $\omega$ brings the spot center along the $z$ axis, in the new position $(0,d)$, being $d$ the distance between the spot and the star centers.}
    \label{fig_rotation}
\end{figure}

The starspot can be found anywhere on the stellar surface. We indicate with $\psi_0$ and $\phi_0$ the latitude and longitude of the spot center. Since we are assuming a spot with circular shape, its projection on the stellar disk will be an ellipse with semi axes $\alpha=r_s$ and $\beta=\alpha \sqrt{1-d^2}$, being $r_s$ and $d$ the radius and the projected distance of the spot from the star center in units of the stellar radius. The spot filling factor is $ff=r_s^2$, while the filling factor projected on the stellar disk is $ff_p=\alpha \beta=r_s^2\sqrt{1-d^2} \leq ff$. According to this approach, for spots at the edge of the star $ff_p=0$.\\
Here we use a coordinate system where $x$-axis is along the line of sight (the observer is at $x=+\infty$), $y$ and $z$ identify the orthogonal plane where the projections are defined and the origin is in the star center (the same convention used in \texttt{pylightcurve}). The projected spot center is in the position $(y_0,z_0)$, where:
\begin{equation}
   \begin{cases}
    y_0 =\cos{\psi_0}\sin{\phi_0}\\
    z_0 =\sin{\psi_0}\\
\end{cases} 
\end{equation}
so the projected distance of the spot is $d=\sqrt{y_0^2+z_0^2}$. The system is invariant under rotations with respect to the $x$ axis and in order to have an analytical solution of the intersections between the $N$ annuli concentric to the star and the elliptical spot we rotate the system around the $x$-axis in such a way the spot center is in along the $z$-axis, in the position $(0,d)$ (see Fig. \ref{fig_rotation}). The rotation matrix $R$ is:
\begin{equation}
R=
\begin{bmatrix}
1 &       0       &      0       \\
0 & \cos{\omega}  & -\sin{\omega}\\
0 & \sin{\omega}  &  \cos{\omega}\\
\end{bmatrix}
\end{equation}
where $\omega$ is the rotation angle (shown in Fig. \ref{fig_rotation}) and it is obtained as: 
\begin{equation}
    \label{rotation_angle}
    \omega=
    \begin{cases}
        \arctan{(\frac{y_0}{z_0})},      & \mbox{if } z_0>0\\
        \sign (y_0) \frac{\pi}{2},       & \mbox{if } z_0=0\\
        -\pi+\arctan{(\frac{y_0}{z_0})}, & \mbox{if } z_0<0\\
    \end{cases}
\end{equation}
The intersection points between the elliptical spot and the $N$ annuli can be found as:
\begin{equation}
\begin{cases} 
    y^2+z^2=r_i^2 \\ 
    \frac{y^2}{\alpha^2}+\frac{(z-d)^2}{\alpha^2(1-d^2)}=1 \\  
\end{cases}
\end{equation}
where $r_i$ is the radius of the $i^{th}$ annulus. After few calculations, we find the solutions $z_{1,2}$:
\begin{equation}
    z_{1,2}=\frac{d\pm d\mu_0\sqrt{\mu_i^2+\alpha^2}}{d^2}=\frac{1\pm\mu_0\sqrt{\mu_i^2+\alpha^2}}{d}
\end{equation}
where $\mu_0=\sqrt{1-d^2}$ and $\mu_i=\sqrt{1-r_i^2}$. For the $y$ coordinates:
\begin{equation}
    y^2 =  \frac{-\mu_0^2-\mu_i^2-\alpha^2\mu_0^2\mp2\mu_0\sqrt{\mu_i^2+\alpha^2}}{d^2}
\end{equation}
Since we want real solutions for $y$, we discard the solution $z=\frac{1+\mu_0\sqrt{\mu_i^2+\alpha^2}}{d}$, so we have two different solutions $y_{1,2}$:
\begin{equation}
    y_{1,2}  = \pm\sqrt{\frac{-\mu_0^2-\mu_i^2-\alpha^2\mu_0^2+2\mu_0\sqrt{\mu_i^2+\alpha^2}}{d^2}}
\end{equation}
To compute the annulus fraction $f(r_i)$ covered by the spot, we distinguish between three cases:
\begin{enumerate}
    \item $d\geq \beta$ (the spot does not include the star center):
    \begin{equation}
        f (r_i)=
        \begin{cases}
            \frac{1}{\pi}\arctan{\left( \frac{y_1}{z}\right)}, &  \mbox{if } |r_i-d|\leq \beta\\
            0, & \mbox{if } |r_i-d|>\beta\\
        \end{cases}
    \end{equation}
    being $y_1$ the positive solution;
    \item $d<\beta$ (the spot occults the origin):
    \begin{equation}
        f(r_i)=
        \begin{cases} 
            1, & \mbox{if } r_i\leq \beta-d\\ 
            \frac{1}{\pi}\arctan{\left( \frac{y_1}{z}\right)}, & \mbox{if } |r_i-\beta| < d \hspace{.2cm}  \mbox{\&} \hspace{.2cm}  z > 0\\ 
            \frac{1}{2}, & \mbox{if }  z = 0\\
            1+\frac{1}{\pi}\arctan{\left( \frac{y_1}{z}\right)}, & \mbox{if } |r_i-\beta|< d \hspace{.2cm}  \mbox{\&} \hspace{.2cm} z< 0\\
            0, & \mbox{if } r_i\geq \beta+d \\ 
        \end{cases}
    \end{equation}
    
    \item $d=0$ (the spot and the star centers coincide):
        \begin{equation}
        f(r_i)=
        \begin{cases} 
            1, & \mbox{if } r_i\leq r_s \\ 
            0, & \mbox{if } r_i> r_s  \\ 
        \end{cases}
    \end{equation}
\end{enumerate}

We remark that each term $f(r_i)$ depends on the spot filling factor and its distance from the star center. In order not to weigh down the notation, we only make explicit the dependence of the radius of $ff$ on $r_i$. We also show that:
\begin{equation}
    \sum_{i=1}^N f (r_i) \cdot 2\pi r_i \cdot \Delta r = \pi ff_{eff}
\end{equation}
being $ff_{eff}=ff \sqrt{1-d^2}$. $ff_{eff}$ is the effective spot filling factor, projected on the star disk.\\
We stress that this mathematical derivation is valid for small spots, in the hypothesis that we can approximate the spot surface with a circle on the plane tangent to the stellar surface in the spot center.
The algorithm presented in this Appendix is implemented in a routine written in \texttt{python}.

\section{Computing the fraction of spotted annulus occulted by the planet}\label{intersection_fraction}
When a planet crosses a spot we have to take into account the flux of the active region occulted by the planet; since the planetary radius $R_\lambda$ is wavelength-dependent, this fraction will depend on the wavelength $\lambda$. By following our annuli integration, we have to compute each fraction $c_\lambda (r_i,t)$ of annulus intercepted by the overlapping area between the spot and the planet at time $t$. We focus on a generic time $t$ where the spot is crossed by the planet. In order to have analytical solutions we rotate both the spot and the planet by the same angle $\omega$ in Eq. \eqref{rotation_angle}. In this work we are interested in analyzing transit systems with high impact parameters, i.e. where the overlapping region between the spot and the planet does not encompass the origin, so we just limit to derive the fraction $c_\lambda (r_i)$ in these cases.\\
The intersection points $(y_{1,2}, z_{1,2})$ between the $N$ annuli and the elliptical spot have been obtained in Appendix \ref{spot_fraction}, while the intersection coordinates $(y_{3,4}, z_{3,4})$ between the $N$ annuli and the planetary external circumference can be found as: 
\begin{equation}
    \begin{cases} 
        y^2+z^2=r_i^2 \\ 
        (y-y_p)^2+(z-z_p)^2= r_\lambda^2 \\  
    \end{cases}
\end{equation}
where $r_\lambda=\frac{R_\lambda}{R_*}$. With a simple algebra we find the intersection points $(y_{3,4}, z_{3,4})$:
\begin{equation}
   \begin{cases} 
    z_{3,4}=\frac{-gd \pm \sqrt{(gd)^2-(g^2+1)(d^2-r_i^2)}}{g^2+1}\\ 
    y_{3,4}=g z_{3,4}+d \\  
\end{cases} 
\end{equation}
being $g=-\frac{z_p}{y_p}$ and $d=\frac{y_p^2 +z_p^2 - r_\lambda^2 + r_i^2}{2 y_p}$. By indicating with $(y', z')$, $(y'', z'')$ the contour points of the overlapping region between the spot and the planet, the fraction $c_\lambda (r_i)$ will be:
\begin{equation}
    c_\lambda (r_i)=\frac{|\arctan{(\frac{y'}{z'})} - \arctan{(\frac{y''}{z''})}|}{2\pi}
\end{equation}

\section{Test of the method developed in Paper I}\label{Discussion}

In this Appendix we discuss the need to take into account the limb darkening effect, by showing the results of the method presented in Paper I and summarized in Section \ref{method}, applied to transit observations simulated with a realistic limb darkening profile. This analysis allows to quantify the influence of the limb darkening effect on the retrieval of the filling factor and temperature of the spot, and of the planet transmission spectrum, so to understand whether or not the limb darkening effect is negligible. In particular, we show the results of the method developed in Paper I to the planetary transits simulated in Section \ref{transit_out_spot}, where for each planet we simulate $n$ transits and the star is simulated with one dominant spot with $\Delta T=1000\,K$ and random size/position on the star disk. At first, we focus on the out-of transit 100 \si{\second} exposures and we perform a bi-dimensional fit where each out-of transit spectrum is compared to a grid of spotted spectra, purposely built for the simulated target, corresponding to different combinations of filling factors and temperatures, in the assumption of uniform emission from the star disk. In order to make a comparison with the method in Paper I, it should be stressed that the filling factors used in Paper I are effective filling factors $ff_{eff}$, i.e. already projected on the stellar disk, therefore the distribution of recovered filling factors from the 2-d fitting procedure is an effective distribution that has to be compared with the projected filling factor of the input spot, depending on its filling factor and distance from the stellar disk center ($ff_{eff}=ff\sqrt{1-d^2}$, see Appendix \ref{spot_fraction}).\\
Fig. \ref{2D_case} shows the distribution of the retrieved spot filling factors and temperatures obtained from the out-of-transit 100 \si{\second} exposures (colored dots) and from the integrated out-of transit observation, for one of the $n$ simulated transits, where the simulated temperature contrast is $\Delta T_{input}=1000\,K$. The colored triangles represent the expected values ($ff_{eff}$, $\Delta T_{input}$),  while the colored crosses are the best-fit spot parameters obtained from the integrated out-of transit observations. For the cases shown in Fig. \ref{2D_case}, it seems that the limb darkening effect can only be neglected for the star HAT-P-11; however, it must be reiterated that the needs to take into account for the limb darkening in the retrieval of the spot parameters is strongly dependent on the position of the spot, i.e. there is an annular region concentric to the star, where the stellar emission is consistent with the hypothesis of a uniform emission from the whole stellar disk. 

\begin{figure}
    \centering
    \includegraphics[width=\columnwidth]{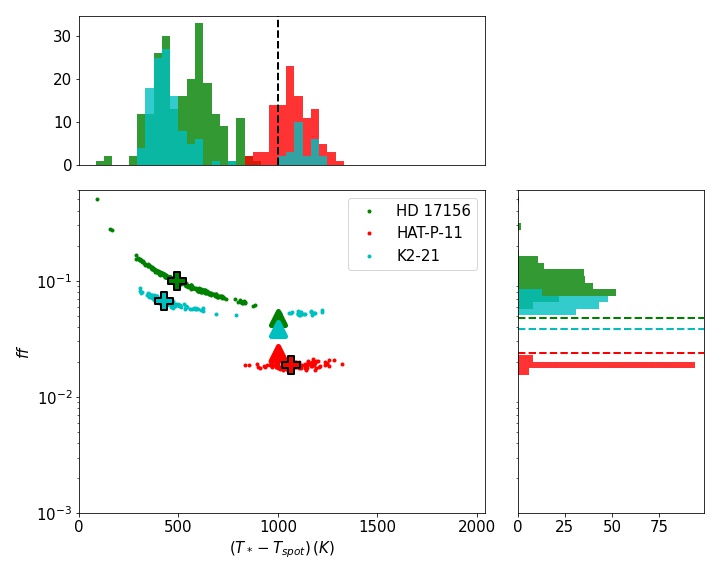}
    \caption{ Scatter plot of the best-fit spot filling factors and temperatures when the star is simulated with a realistic limb darkening profile and the resulting spectrum is fitted by neglecting the limb darkening effect. The values have been obtained from the 100 \si{\sec} out-of transit exposures in one of the $n$ simulated transits and each color pertains to a specific star (reported in the legend). The triangles are the projected filling factor and temperature contrast of the simulated spot, while the colored crosses are the best-fit values obtained when summing all the out-of transit exposures. The upper histogram shows the distribution vs. derived $\Delta T$, the one on the right the analogous distribution vs. derived effective $ff$. The dashed lines in each histogram mark the expected values.}
    \label{2D_case}
\end{figure}

\begin{figure}
    \centering
    \includegraphics[width=\columnwidth]{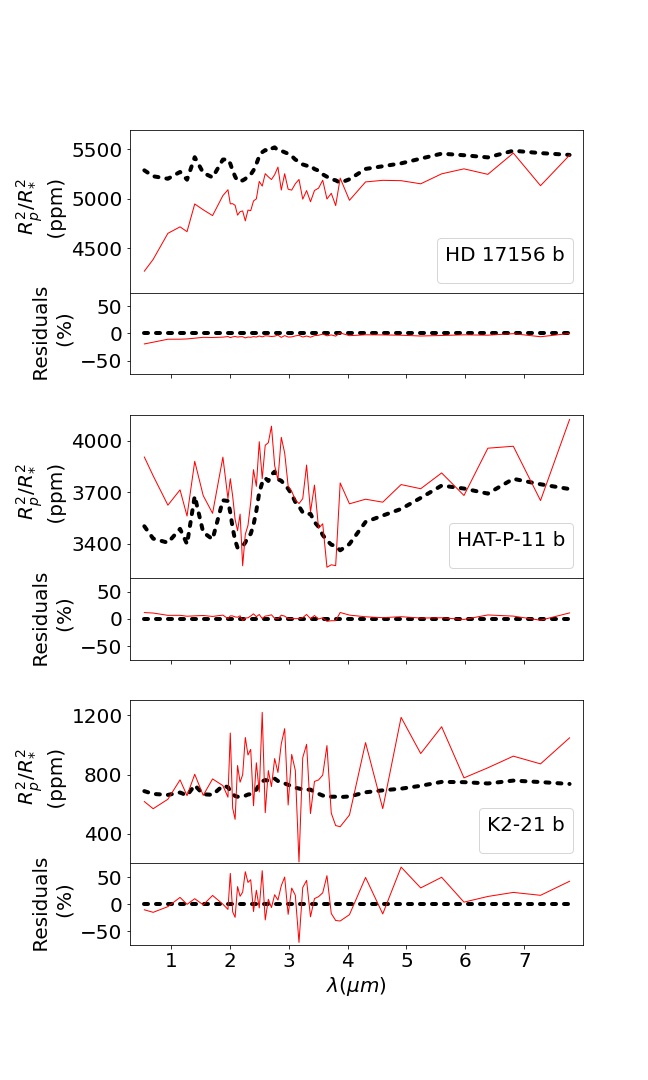}
    \caption{Retrieved transmission spectra when applying the method described in Paper I to the planetary transits shown in Fig.\ref{retrieved_pl_spectra}.}
    \label{2D_case_spectra}
\end{figure}

Following the approach described in Paper I, for each of the $n$ simulated transits we consider the best-fit spot parameters derived from the integrated out-of transit observation, and from these we derive: at first, the spectrum of the unspotted photosphere $P_\lambda(T_*)$, then the planet spectrum as $\frac{F_\lambda^{out}-F_\lambda^{in}}{P_\lambda(T_*)}$, being $F_\lambda^{out}$ and $F_\lambda^{in}$ the out-of transit and the in-transit observed fluxes, respectively (by neglecting the egress and ingress of the planet from the transit). For each planetary system, we average the transmission spectra thus obtained in the $n$ transits. Fig. \ref{2D_case_spectra} shows the retrieved transmission spectra for the three planets: in all the explored cases, the retrieved spectra (red continuous lines) are systematically distorted if compared to the input spectra (black dashed lines). These results are not due to the spot as its influence is removed through the difference $F_\lambda^{out}-F_\lambda^{in}$, but rather to the fact that the star does not have a uniform emission, i.e. the  emission in the occulted portion of star is overestimated (HAT-P-11 b, K2-21 b) or underestimated (HD 17156 b) with respect the averaged emission from the whole stellar disk. Therefore, this effect is strongly dependent on the impact parameter of the transiting planet.

\bsp	
\label{lastpage}
\end{document}